\documentclass[aps,prb,preprint,groupedaddress,showpacs,showkeys]{revtex4-1}
\usepackage{graphicx}
\usepackage{threeparttable}

\begin{document}

\title{Electrical Transport Properties of Co-based Skutterudites filled with Ag and Au}

\author{Maria Stoica}
\author{Cynthia S. Lo}
\email[]{clo@wustl.edu}
\affiliation{Department of Energy, Environmental and Chemical Engineering, Washington University, 1 Brookings Drive, Saint Louis, Missouri  63130, USA}

\date{\today}

\begin{abstract}
This work presents theoretical calculations of the electrical transport properties of the Ag, Au, and La fractionally filled bulk skutterudites: CoSb$_{3}$, CoAs$_{3}$, and CoP$_{3}$. Density functional theory (DFT), along with projector augmented wave (PAW) potentials, was used to calculate bulk band structures and partial density of states. The Seebeck coefficient ($S$), electrical conductivity ($\sigma$), and power factor ($S^2 \sigma$) were calculated as a function of temperature and filling fraction using the momentum matrix method along the entire first Brillouin zone. Calculated trends in the electrical transport properties agree with previously published experimental measurements for p-type unfilled and La filled CoSb$_3$. The calculated $S$ and $\sigma$ values for the Ag and Au filled compounds indicates that the most promising electronic properties are exhibited by p-type Au$_{0.125}$(CoSb$_{3}$)$_{4}$, Au$_{0.25}$(CoSb$_{3}$)$_{4}$, and Au(CoSb$_{3}$)$_{4}$. Au is therefore recommended as a promising filler for improved thermoelectric properties in cobalt antimonides. Ag is also a good filler for cobalt phosphides; the creation of a negative indirect band gap is observed in Ag(CoP$_3$)$_4$, which indicates semimetallic behavior, so this compound may possibly exhibit lower thermal conductivity than metallic CoP$_3$. Finally, we recommend future directions for improving the thermoelectric figure of merit of these materials.
\end{abstract}

\pacs{72.20.Pa, 71.20.Nr, 71.15.Mb}

\keywords{density functional theory, filled skutterudite, thermoelectric power factor, electrical transport properties}

\maketitle

\section{\label{sec_intro}Introduction}

Thermoelectric materials can be used for solid state power generation in a variety of applications, including automotive power generation \cite{Bell_Science_ThermoelectricApps,Yang_IEEEroc_ThermoelectricsAppsAuto} and solar energy capture and conversion \cite{Tritt_MRSBul_ThermoelectricSolar}. Such devices operate without the need for moving parts, and can be incorporated into most current energy generation technologies to capture waste heat and increase conversion efficiency. Current thermoelectric devices are relatively inefficient ($\approx$ 18\% \cite{Tritt_MRSBul_ThermoelectricPhenomenaApps}) but with recent advances in nanostructuring concepts, it has been shown that more efficient devices may be attainable \cite{Mahan23071996}. 

\begin{figure}
\includegraphics[width=130mm]{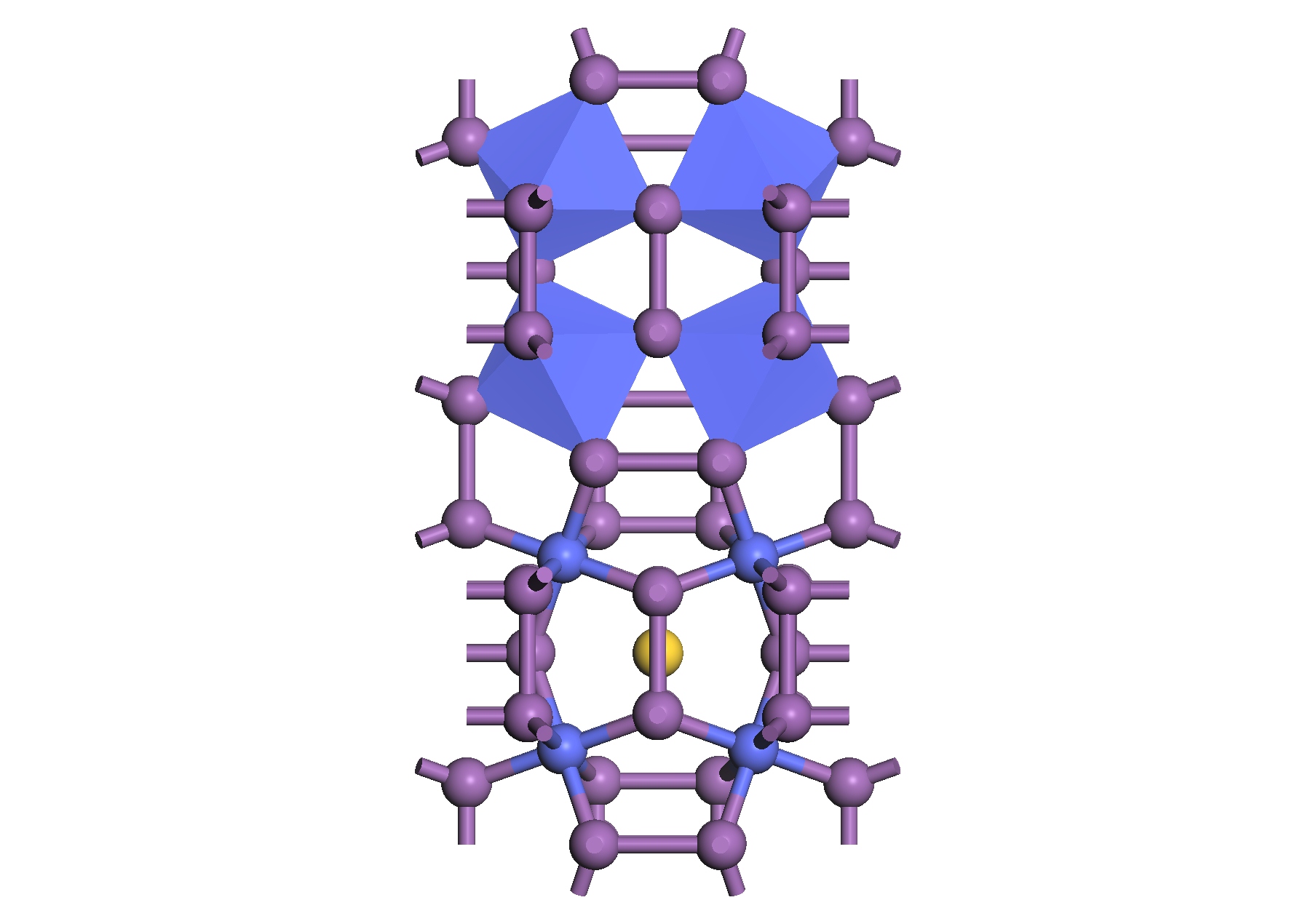}
\caption{Skutterudite (CoM$_3$) unit cell structure. The top unit cell shows the sixfold coordinated Co atoms (blue). The bottom unit cell shows a filler atom (gold) inserted in the center void. The second void in the unit cell is located at the corners.\label{fig1_skutterudite_structure}}
\end{figure}

In this study, we focus specifically on skutterudite materials, which are attractive for thermoelectric applications because of their large unit cells and high intrinsic electrical properties. The skutterudite unit cell consists of eight TM$_{6}$ octahedral groups -- one in each corner of the cubic unit cell -- where T is a transition metal (e.g., Co, Rh, or Ir) and M is a metalloid or nonmetal (e.g., Sb, As, or P) (Figure \ref{fig1_skutterudite_structure}). Each M atom belongs to two adjacent octahedra and each T atom is at the center of the octahedral group. These octahedra are arranged in such a way that a cage-like void is created in the center of the unit cell. These voids are large enough to incorporate certain atoms (X) that, when present, have the ability to lower the thermal conductivity of the material by "rattling" about within the void, decreasing the phonon mean free path, and thus, increasing phonon dispersion. 

Previous studies have shown that filling the naturally occurring voids within the skutterudite lattice with heavy element atoms, (e.g., Lanthanides La, Ce, and Eu), leads to a considerable reduction in the thermal conductivity \cite{Sales_PRB_LaCeFilledCoSb3,Lamberton_APL_EuFilledCoSb3}. In general, ideal filler atoms are described as small and heavy, with electrons that can readily hybridize with the conducting electrons in the lattice \cite{PhysRevB.53.1103,Nolas_MRSBul_RecentDevelThermoelectrics}. These hybridized electrons may help prevent large reductions in electronic properties that might otherwise occur through void filling. Recent studies have indicated that the filling of the lattice voids with smaller or lighter atoms with more metallic properties, such as Ba \cite{Chen_JAP_BaFilledCoSb3}, Na \cite{Pei_APL_NaFilledCoSb3} or In \cite{He_ChemMat_InFilledCoSb3}, leads to an increase in the thermoelectric figure of merit compared to that obtained by filling with La. This is in part attributable to the enhancement of electronic properties from metallic fillers as compared to heavier Lanthanides. Furthermore, it has been shown that double element filling may further enhance thermoelectric performance \cite{Yang_APL_DoubleFilledCoSb3} compared to single element filling, though this is outside the scope of the present work. 

In this study, we test the accuracy of density functional theory (DFT) calculations, following the Projector Augmented Wave (PAW) approach, for predicting electrical transport properties for skutterudite thermoelectric materials; we use the momentum matrix method in conjunction with PAW (PAW-MM) and compare our calculated results to previously published experimental results. Many previous theoretical investigations into skutterudites have utilized the full-potential linearized augmented plane wave (FLAPW) approach to calculate electronic properties, but we are specifically interested in PAW for its computational speed and efficiency. These qualities are desirable when screening a large group of prospective materials for a specific application. We calculate the thermoelectric properties for both well-characterized and previously uncharacterized skutterudites.  We emphasize that the computational method employed is a simplified calculation for ideal bulk crystal systems that does not take into account grain boundary effects or the dependence of the electronic time relaxation constant, $\tau_e$, on temperature. Therefore, only trends and relative magnitudes of properties are demonstrated here. 

We also aim to determine the effect of filler size, atomic weight, and electronegativity on the resulting electrical transport properties of filled Co-based skutterudites. A summary of filler properties is presented in Table \ref{table_element_breakdown}. We hypothesize that Au is a promising metallic filler atom due to its small size and large mass, and because its valence electron configuration is similar to that of La, which has been previously shown to be a successful filler. Both Au and La have similar core electronic structures and valence shells composed of all but one filled orbitals and one unpaired electron in the highest energy orbital; they differ only in that the unpaired electron in La is a $d$ electron and that in Au is an $s$ electron. Furthermore, Au is heavier (atomic weight is 197 amu) and smaller (atomic radius is 135 pm) than La (atomic weight is 139 amu, atomic radius is 195 pm). Both of these desirable characteristics should result in improved thermoelectric properties for Au filled skutterudites. We also study Ag as a candidate filler, since Ag has a similar electronic structure compared to Au, but is lighter (atomic weight is 108 amu) and larger (atomic radius is 144 pm). From size considerations, we would expect that the size of the filler atom to be important if the atom is large enough to interact with the void boundaries; otherwise, the heaviness of the atom is likely to be more important for phonon dispersion.  Finally, we note that the electronegativities of Ag, Au, and La are 1.9, 2.4, and 1.1, respectively, and those of P, As, and Sb are 2.1, 2.0, and 1.9, respectively. We expect that larger differences in electronegativity between the filler and host (M) atoms would improve electron transport in filled skutterudites. All of these effects will be explored in the calculation of the thermoelectric strength of Ag, Au, and La filled Co-based skutterudites.
\section{\label{sec_methods}Methods}

\subsection{\label{subsec_theory}Theory}

One quantity that is commonly used to characterize the thermoelectric strength of a material is the thermoelectric figure of merit  \cite{Tritt_MRSBul_ThermoelectricPhenomenaApps}, $z$: 
\begin{eqnarray}
\label{zT_material}
zT & = & \frac{S^2 \sigma}{\kappa} T = \frac{S^2}{\rho \kappa}.
\end{eqnarray}
Here, $S$ is the Seebeck coefficient, $\sigma$ is the electrical conductivity (with $\rho$ being its inverse, or the electrical resistivity), and $\kappa$ is the thermal conductivity. We aim to maximize $S$ and $\sigma$ and minimize $\kappa$ for optimal thermoelectric performance. Equation \ref{zT_material} suggests that for every increase in the Seebeck coefficient, $\Delta S$, there is a corresponding increase, $\left( \Delta S \right)^2$, in $zT$. $\sigma$ and $\kappa$ are directly and inversely proportional to $zT$, respectively. In this work, we will focus on optimizing the electrical transport properties of skutterudites by characterizing $S$ and $\sigma$.

Traditionally, band structure calculations have been used to directly calculate electronic properties. However, the problem with this approach is that many DFT software packages output the band structure in the form of an array organized by increasing energy, without band distinction. Thus, errors may arise in the computation of the electrical conductivity, particularly at band crossings around the Fermi level, even when Fourier transformations are used \cite{Madsen_CPC_Boltztrapp}. A different approach is desirable in complex systems such as skutterudites, when numerous band crossings are present. 

The momentum matrix method rectifies this issue by relying directly on the wavefunction descriptors. We can express the computed electron group velocities as follows\cite{Scheidemantel_PRB_TransportCoeffsFirstPrinciples}:
\begin{eqnarray} 
\label{momentum_method}
v_{\alpha\beta} \left( i,k \right) & = & \frac{1}{m} \langle \psi_{\alpha\beta} \left( i,k \right) \left| \hat{p} \right| \psi_{\alpha\beta} \left( i,k \right) \rangle
\end{eqnarray}
Here, $v$ are the velocities, $\psi$ are the wavefunctions, $m$ is the electron mass, and $\hat{p}$ is the momentum operator. $\left( i,k \right)$ denotes the eigenvalue-band index and $\alpha\beta$ denote the directional coordinates. The resulting matrix of velocities indexed by energy and $k$-point is called the optical matrix. Although this equation may be used to determine group velocities from exact wavefunction descriptions, the PAW-MM approach requires the addition of an augmentation factor to account for errors arising from the pseudization of the wavefunctions around the atomic cores\cite{Furthmuller_PRB_OpticalPropertiesUsingPAW,Kageshima_PRB_MomMatrixCalcPP}. 

Boltzmann transport theory is then used to derive the following nested equations that enable us to determine $\sigma$ and $S$ in terms of the group velocities\cite{Madsen_CPC_Boltztrapp}:
\begin{eqnarray}
\label{eqs_properties1} \sigma_{\alpha\beta} \left( i,k \right) & = & e^2 \tau_{i,k} v_\alpha \left( i,k \right) v_\beta \left( i,k \right) \\
\label{eqs_properties2} \sigma_{\alpha\beta} \left( \varepsilon \right) & = & \frac{1}{N} \sum_{i,k} \sigma_{\alpha\beta} \left( i,k \right)\frac{\delta \left( \varepsilon-\varepsilon_{i,k} \right)}{d\varepsilon} \\
\label{eqs_properties3} \sigma_{\alpha\beta} \left( T; \mu \right) & = & \frac{1}{\Omega} \int \sigma_{\alpha\beta} \left( \varepsilon \right) \left[ -\frac{\partial f_\mu \left(T;\varepsilon \right)}{\partial\varepsilon} \right] d\varepsilon \\
\label{eqs_properties4} \nu_{\alpha\beta} \left( T; \mu \right) & = & \frac{1}{eT\Omega} \int \sigma_{\alpha\beta} \left( \varepsilon \right) \left( \varepsilon-\mu \right) \left[ -\frac{\partial f_\mu \left( T; \varepsilon \right)}{\partial\varepsilon}\right] d\varepsilon   \\
\label{eqs_properties5} S & = & \frac{\nu_{\alpha\beta} \left( T; \mu \right)}{\sigma_{\alpha\beta} \left( T; \mu \right)} 
\end{eqnarray}
Here, $N$ is the number of $k$-points, $\delta$ is the unit impulse function, $\Omega$ is the volume of the primitive cell, $\varepsilon_F$ is the Fermi level, and $f$ is the Fermi-Dirac distribution for an electron gas:

\begin{eqnarray}
f_0 \left( \varepsilon \right) & = & \left[ \exp \left( \frac{\varepsilon - \varepsilon_{F}}{k_B T} \right) + 1 \right]^{-1}
\end{eqnarray}

Equations \ref{eqs_properties3}-\ref{eqs_properties5} yield second-rank tensors. To quantify these properties as scalars, the trace of each matrix is evaluated.

\subsection{Calculations}

The electrical transport properties of fractionally filled Co-based skutterudites -- specifically, CoSb$_{3}$, CoAs$_{3}$, and CoP$_{3}$ with Ag, Au, and La filler atoms -- were evaluated. La was chosen for its previously demonstrated ability to increase the thermoelectric figure of merit of filled skutterudites; thus, La will serve as a baseline for comparison to the previously uncharacterized Ag and Au fillers.

All calculations of electronic wavefunctions were performed using DFT, as implemented in the Vienna Ab Initio Simulation Package (VASP) \cite{Kresse_CompMatSci_VASPRef0,Kresse_PRB_VASPRef1,Kresse_PRB_VASPRef3,Kresse_CompMatSci_VASPRef2}. PAW potentials \cite{Bloechl_PRB_PAWMethod,Kresse_PRB_PSSToPAW} were used to maintain a balance between high accuracy and reasonable computational costs. The generalized gradient approximation of Perdew, Burke, and Ernzerhof \cite{Perdew_PRLet_PBEGGA,Perdew_PRLet_PBEGGAErrata} was used to represent the exchange-correlation energy. All calculations were performed using the blocked Davidson iterative matrix diagonalization algorithm\cite{davidson_methcomputmolphys_1983} to optimize the orbitals. The tetrahedron method with Bl\"{o}chl corrections was used to determine partial occupancies. The first Brillouin zone was sampled with 100 $k$-points, and the kinetic energy cutoff for the plane wave basis set was set to 270eV after extensive $k$-point convergence tests. The resulting wavefunctions are used in the momentum matrix equation \ref{momentum_method}, with a PAW potential correction to determine the group velocities as noted in Section \ref{subsec_theory}. Equations \ref{eqs_properties1}-\ref{eqs_properties5} are then used to determine the thermoelectric properties from the group velocities with a code developed in MATLAB \cite{matlab}. 

As shown in Figure \ref{fig1_skutterudite_structure} there are two voids per unit cell, so that only a limited number of filling fractions ($x \in \left\{ 0.5, 1 \right\}$) may be explored. In order to obtain a larger sampling set of filling fractions, a $2 \times 2 \times 2$ supercell containing 16 voids was constructed. Voids were filled in pairs until the structure was fully filled. Table \ref{table_symmetries} shows the resulting highest symmetries used for the filled systems,X$_x$(CoM$_3$)$_4$. $S$, $\sigma$, and $S^2 \sigma$ values were calculated for $x \in \left\{ 0.125, 0.25, 0.375, 0.5, 0.625, 0.75, 0.875, 1 \right\}$; a denser sampling of filling fractions would be difficult to achieve given computational limitations and may not yield significantly different trends in the resulting data.  All calculations were performed at the following temperatures: $T \in \left\{ 300, 400, 500, 600, 700, 800, 900 \right\} \text{ K}$.

Calculations were first performed on the unfilled CoSb$_3$ system and the La filled, Fe doped CoSb$_3$ system at different carrier concentrations, and compared to previously published experimental data. Doping was mimicked using the rigid band model (i.e., moving the Fermi energy on a fixed band structure) to simulate different electron carrier concentrations, $n_e$, and hole carrier concentrations, $n_h$. These carrier concentrations were determined using the density of states, $g \left( \varepsilon \right)$:

\begin{eqnarray}
n_e & = & \frac{1}{V} \sum\limits_{\varepsilon=\varepsilon_F}^{\infty}g_c \left( \varepsilon \right) f \left( \varepsilon \right) \Delta\varepsilon\\
n_h & = & \frac{1}{V}\sum\limits_{\varepsilon \rightarrow -\infty}^{\varepsilon_F} g_v \left( \varepsilon \right) \left( 1-f \left( \varepsilon \right) \right) \Delta\varepsilon
\end{eqnarray}

The $c$ and $v$ subscripts denote the conduction and valence states respectively. Then, the Ag, Au, and La filled (undoped) skutterudites were screened across the range of $x$ and $T$ values in the sample space, and the electrical transport properties -- $S$, $\sigma$, and $S^2\sigma$ -- were calculated from the wave descriptors. Additional calculations of the electrical transport properties were performed on the most promising filled systems for a range of deviations in the chemical potential.

\section{\label{sec_results}Results and Discussion}

\subsection{Unfilled skutterudites}

The rightmost column in Table \ref{table_lattice} shows the calculated void radii as determined upon geometry optimization of the skutterudite systems. Full lattice optimizations resulted in only a +0.6\% increase in the lattice parameters; thus, we allowed only the atoms to relax during geometry optimization. Smaller voids are observed in CoAs$_{3}$ and CoSb$_{3}$ and larger voids are observed in CoP$_{3}$ upon geometry optimization. 

\begin{figure}
\includegraphics[width=130mm]{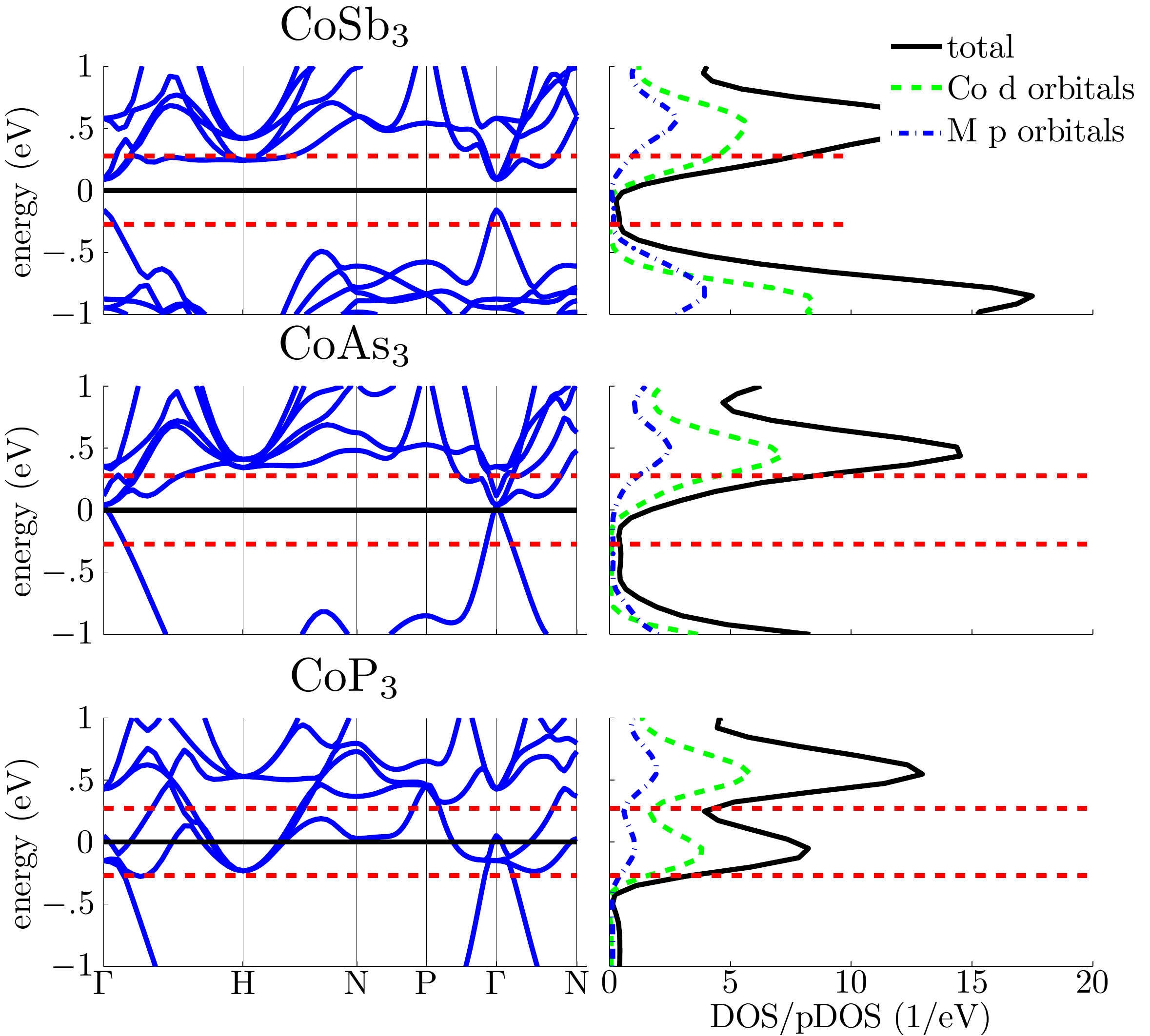}
\caption{Band structure and partial density of states for unfilled skutterudites.\label{fig2_bsdos_unfilled}}
\end{figure}

Figure \ref{fig2_bsdos_unfilled} shows the band structures calculated for the three unfilled skutterudites (i.e., CoP$_{3}$, CoAs$_{3}$, and CoSb$_{3}$). CoSb$_3$ is a semiconductor with a small band gap of 0.3 eV, CoAs$_3$ has a pseudo gap of 0 eV, and CoP$_3$ is a conductor with an overlap at the $\Gamma$ point; these results generally agree with previous findings \cite{PhysRevB.58.15620,PhysRevB.71.155119}. The total density of states (DOS), presented alongside the partial contributions from the Co 3d electrons and the 3p, 4p, and 5p electrons of P, As, and Sb, respectively, are also shown in Figure \ref{fig2_bsdos_unfilled}. The dotted horizontal lines denote the Fermi-Dirac distribution filter ($\frac{\partial f}{\partial \varepsilon}$) of equations \ref{eqs_properties3}-\ref{eqs_properties4} at 300K in this and all subsequent figures.

From these plots we observe that there is significant hybridization of the valence orbitals at the Fermi level for all three bulk materials. We also notice that a majority of the DOS contribution around the Fermi energy may be attributed to the valence electrons. These DOS and band structure diagrams indicate that the valence electrons are essential in dictating the electronic properties of the compound, and fillers containing electrons that may hybridize with electrons at these energy levels can strongly impact electrical transport properties.

\subsection{\label{skut_exp}Comparison of Computational to Experimental Results}

\begin{figure}
\includegraphics[width=150mm]{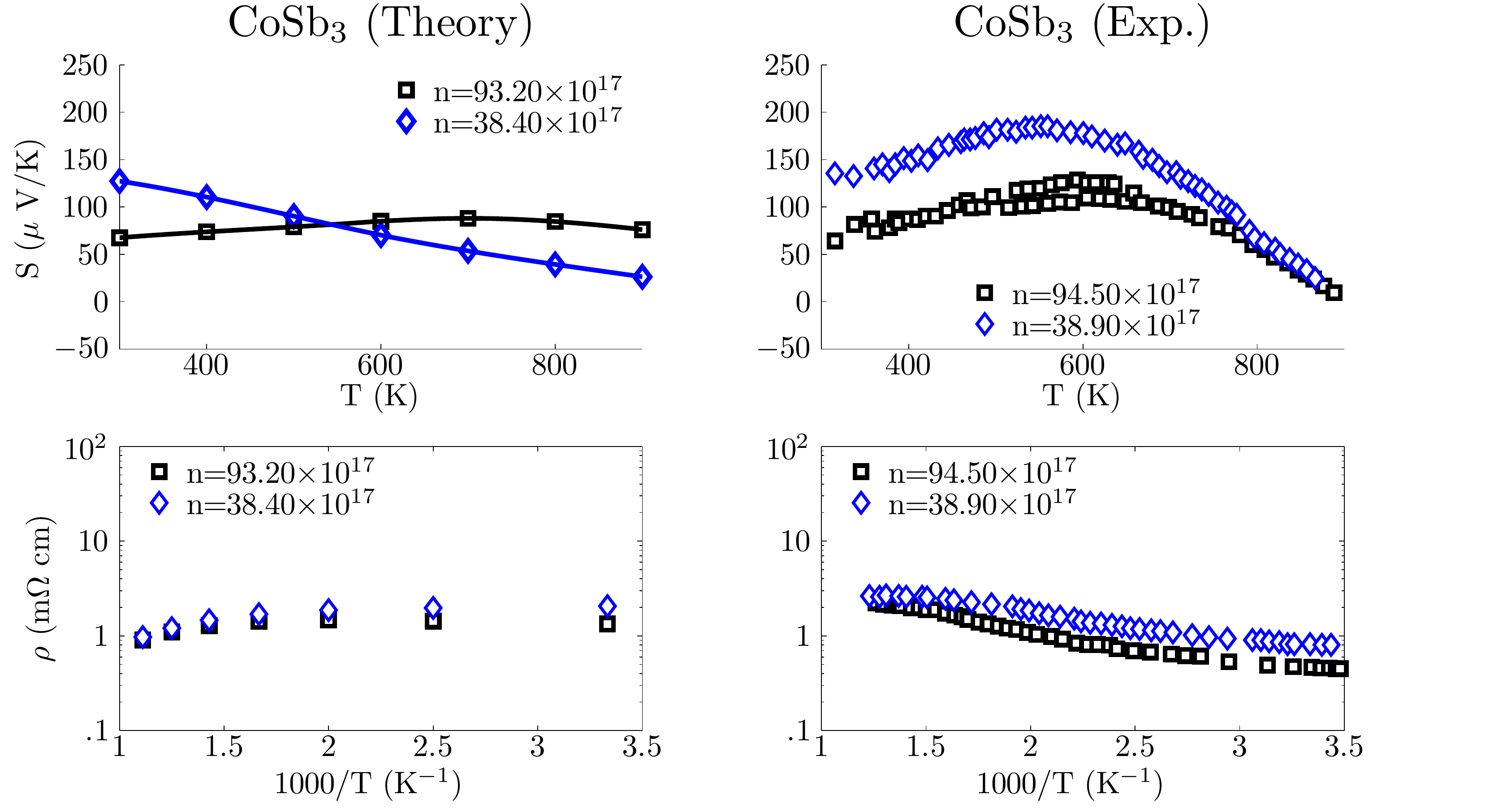}
\caption{Comparison of theoretically computed and experimental \cite{Caillat_JAP_PropertiesSingleCrystallineCoSb3} electrical transport properties, $S$ and $\rho$, for unfilled p-type CoSb$_3$, at different hole carrier concentrations, $n$.\label{fig3_theoretical_v_exp_unfilled}}
\end{figure}

We first compare the calculated Seebeck coefficient, $S$, and electrical resistivity, $\rho$, to available experimental values for two systems -- unfilled CoSb$_3$ and La filled, Fe doped CoSb$_3$. Figure \ref{fig3_theoretical_v_exp_unfilled} shows that the trends in both properties and the peak in $S$ are predicted correctly for unfilled p-type CoSb$_3$ at high hole carrier concentrations. By comparing our calculated values of $\rho$ to the experimental data at $T = 500 \ \text{K}$, we determine that the best-fit value for the electronic time relaxation constant for unfilled CoSb$_3$ is $\tau_e=5\times 10^{-13} \ \text{s}$. For low carrier concentrations, (i.e., where the Fermi energy approaches the undoped value), our theoretical values of $S$ drop off more rapidly than the observed experimental values. At room temperature, our calculated $S$ values are correct; however, as the temperature rises, the discrepancy increases. Since we determine our band occupations from the Fermi distribution, it is possible that the deviations arise from the inclusion of conduction bands in the calculation. As the temperature increases, more conduction bands are included in the computation. Furthermore, as the hole carrier concentration increases ($\varepsilon_F$ moves downward on the band structure), the Fermi distribution includes fewer conduction electrons and our estimation improves. This is a reflection of the limitations of DFT in describing conduction band characteristics, especially at band edges. Although the theoretically predicted and experimentally measured values of $\rho$ and $S$ do not always agree throughout the sampled temperature range, the trends in these properties with temperature are consistent and demonstrate that there is an optimal temperature range for high values of $S$ and low values of $\rho$.

\begin{figure}
\includegraphics[width=150mm]{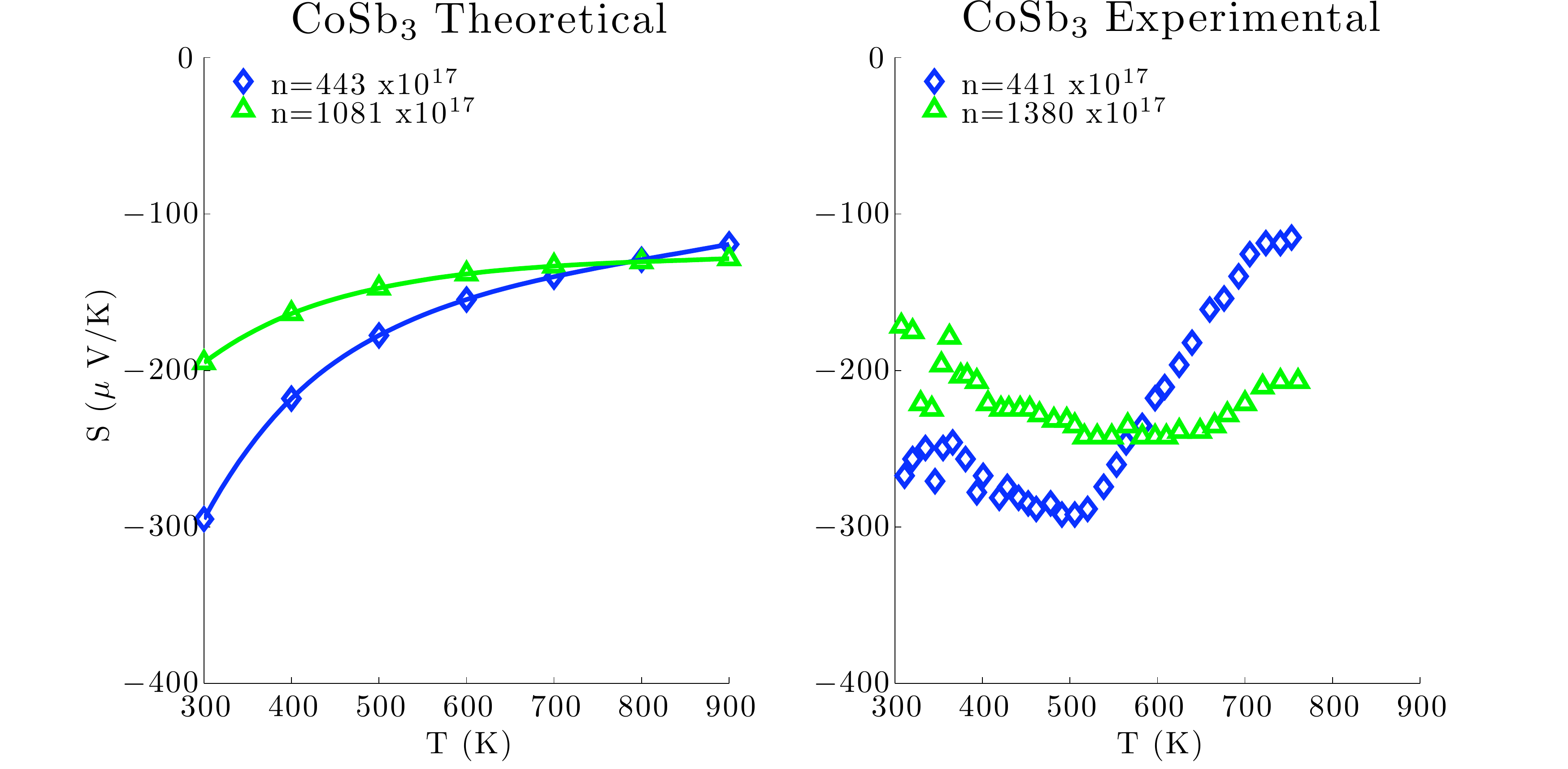}
\caption{Comparison of theoretically computed and experimental \cite{PhysRevB.72.085126} electrical transport property, $S$, for unfilled n-type CoSb$_3$, at different hole carrier concentrations, $n$.\label{fig_ntype_comparison}}
\end{figure}

Although not directly the focus on this work, we note that that there have been prior comparisons made between theory and experiment for unfilled n-type CoSb$_3$ \cite{0953-8984-15-29-315,PhysRevB.72.085126}, so we now demonstrate that our model is flexible enough to consider both types of dopants.  For n-type materials, the number of bands included in the calculation must be increased in order to ensure convergence of the behavior of the conduction bands. Figure \ref{fig_ntype_comparison} shows a comparison between the experimental trends and the theoretical trends for unfilled n-type CoSb$_{3}$. For high electron carrier concentrations, where the Fermi energy lies within the conduction band, we achieve good agreement between theory and experiment. The carrier concentration used in the theoretical calculations was obtained by fitting the computed $S$ to the experimental data at $T = 300 \ \text{K}$. As depicted in Figure \ref{fig_ntype_comparison}, the shape of the computed curves does not exactly match the experimental data across the range of temperatures sampled, although the computed values of $S$ at low temperatures match within 10\% of the experimental values. We attribute this discrepancy to the observation that in experimentally developed materials, the carrier concentration and transport properties will fluctuate differently with temperature, as a result of defects, than what would be expected in a defect-free model like that used in our calculations.  Deviations in the steepness of the curvature of the bands are thus most likely due to the constant relaxation time estimation \cite{PhysRevB.72.085126} used in our calculations.  For very low carrier concentrations (not shown), where the chemical potential falls within the band gap, we observe a greater deviation from the experimental results, especially at low temperatures. Since the Fermi-Dirac distribution does not accurately describe the distribution of electrons within the band gap, we must instead increase the k-point sampling of the Brillouin zone to 5,000 points and triple the number of total bands in the calculation to achieve reasonable accuracy within the band gap. However, we again stress that for the filled systems considered subsequently in this study, the filler acts as a p-type dopant and the Fermi level is pushed into the valence band, so that only heavily doped materials are observed; therefore, such resource-intensive calculations on n-type systems are outside the scope of the present work.

\begin{figure}
\includegraphics[width=150mm]{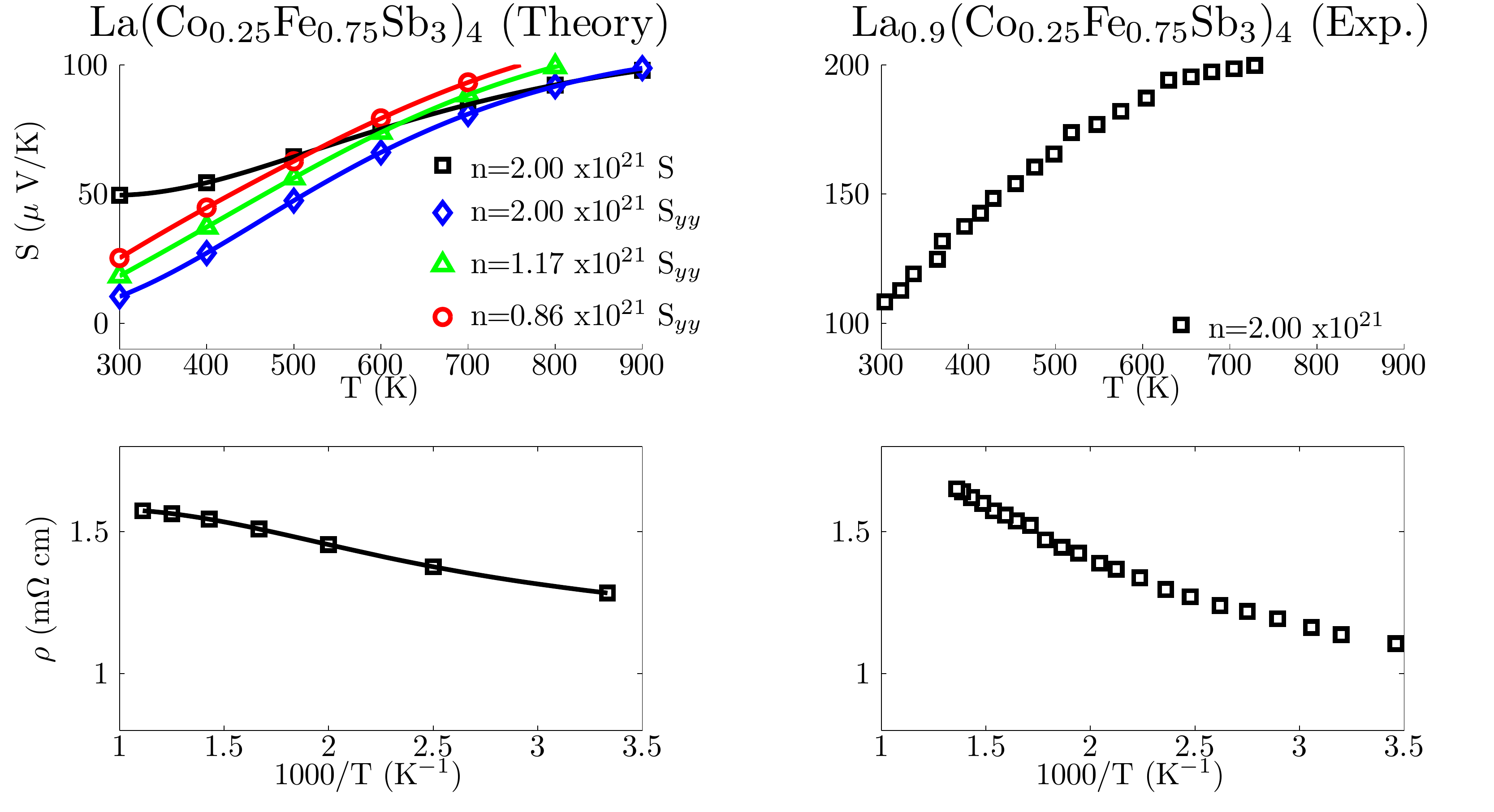}
\caption{Comparison of theoretically computed and experimental \cite{Sales_PRB_LaCeFilledCoSb3} electrical transport properties, $S$ and $\rho$, for La(Co$_{0.25}$Fe$_{0.75}$Sb$_3$)$_4$ and La$_{0.9}$(Co$_{0.25}$Fe$_{0.75}$Sb$_{3}$)$_4$, respectively.\label{fig5_theoretical_v_exp_la}}
\end{figure}

The results for La filled, Fe doped skutterudites are shown in Figure \ref{fig5_theoretical_v_exp_la}. The experimental measurements are made on 90\% La filled skutterudites, but given the range of filling fractions that were computationally tractable, we relate these results to our calculated predictions on 100\% La filled skutterudites. We see from these plots that the trace  of the Seebeck coefficient has a flatter temperature dependence compared to the experimental trends. However, after investigating the directional components of $S$, we see that the $S_{yy}$ component exhibits a similar trend to the experimental values. As we adjust the value of $n$ to attempt to emulate 90\% La filling, we notice that the values asymptotically approach the experimental results. Thus we conclude that the small discrepancy between the calculated and experimental values of $S$ is likely due to the difference in the La filling fraction.  Again, we fit the calculated values of $\rho$ to the experimental data at $T = 500 \ \text{K}$, and determine that the best-fit value for the electronic time relaxation constant for La filled, Fe doped CoSb$_3$ is $\tau_e=2 \times 10^{-13} \ \text{s}$. From these results, it seems reasonable to estimate the electronic relaxation time constant for all of the systems considered in this study to be $\tau_e= 10^{-13} \ \text{s}$, and we will use this value for all subsequent calculations.

\subsection{\label{skut_symm}Ag and Au filled skutterudites}

\begin{figure}
\includegraphics[width=130mm]{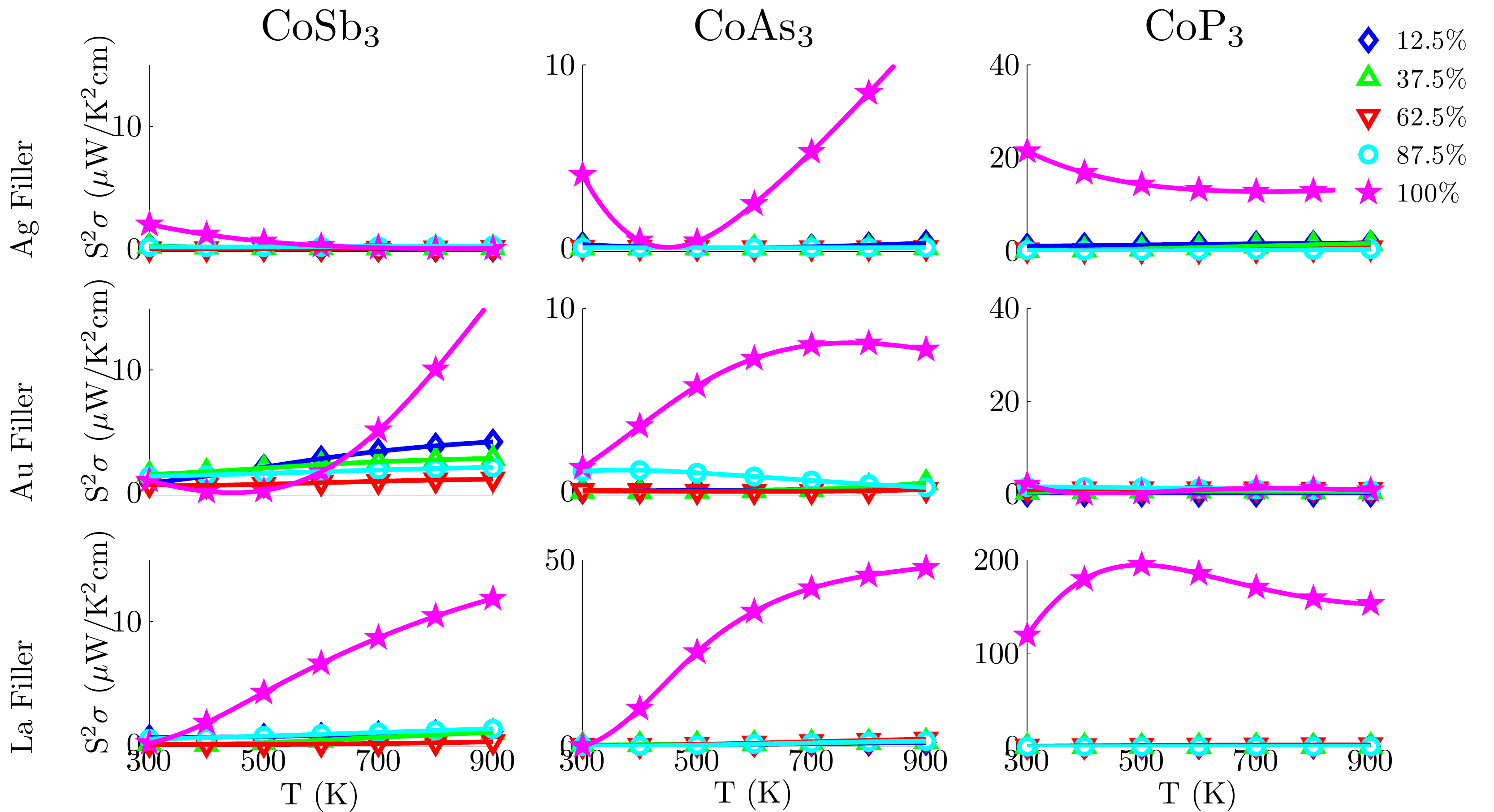}
\caption{Comparison of Ag, Au, and La fillers for CoM$_3$ (M = P, As, Sb) skutterudites at filling fractions corresponding to Im$\overline{3}$ symmetry in the unit cell.\label{fig6_powerfactors_all}}
\end{figure}

We next explored Ag and Au filled skutterudites as potential candidates for improved thermoelectric behavior.  We plotted the calculated power factor, S$^{2}\sigma$, which is the numerator of the figure of merit, as a function of temperature. For these calculations, the time relaxation constant was again assumed to be independent of filling fraction and temperature, and was approximated as $\tau_e=10^{-13} \text{ s}$, as described in Section \ref{skut_exp}. Figure \ref{fig6_powerfactors_all} shows the power factors plotted as a function of temperature for the undoped compounds at the filling fractions that yielded Im$\bar{3}$ symmetry of the unit cell. Charge compensation was not included so that the effect of composition on the thermoelectric power factor may be isolated and identified as follows:

\begin{itemize}
\item CoP$_3$: Ag is a reasonably good filler at 100\% filling fraction above room temperature. Au, however, is not a good filler in this compound. La is an excellent filler and actually enhances electronic properties for fully filled lattices.
\item CoAs$_3$: Both Au and Ag are good fillers at temperatures above 500 K. The power factor for Ag(CoAs$_3$)$_4$ experiences a dip at 450 K because of the transition from p-type to n-type behavior, as reflected by the sign of $S$. La is a moderately good filler. However, completely filling the lattice with Au seems to have an enhanced effect on electrical transport properties compared to partial filling, so this system is worthwhile as the subject of future research for high temperature applications.
\item CoSb$_3$: Ag is a poor filler. Au holds great promise as a filler, especially for low filling fractions at temperatures between 500 K and 800 K and for complete filling at temperatures above 700 K. Indeed, complete filling with Au exhibits the greatest enhancement of electrical transport properties at very high temperatures, even when compared with La, which is an excellent filler in fully filled systems.
\end{itemize}

Figure \ref{fig6_powerfactors_all} indicates the potential of fillers for high-performing thermoelectric behavior, based on undoped systems. The power factors at 100\% filling are markedly higher because of a multiple-fold increase in the electrical conductivity, specifically in the $\sigma_{zz}$ component. This is likely due to increased coordination between the filler atoms and the host lattice in the fully filled configuration. Ag and Au are excellent conductors, and in a fully filled skutterudite lattice, the layout of the fillers is similar to that in the pure conductors. Therefore, when the voids are fully filled, there may be resonance at the location of the fillers, so that a highly electrically conductive material would be produced. Experimentally, this is corroborated by 90\% La filled CoSb$_3$ exhibiting the best performance among similar skutterudites \cite{Sales_PRB_LaCeFilledCoSb3}.

Although the fully filled systems in general show the greatest promise, 100\% filling is rarely achievable without heavy doping for charge compensation. Furthermore, a system that does not possess high power factors when undoped may still be a good thermoelectric when doped, since it is also likely that fully filled materials would exhibit high thermal conductivities without the introduction of impurities or defects. In any event, screening calculations on undoped systems, as performed here, can aid in highlighting systems where the filler acts as a dopant and enhances electrical transport properties.  For the examples of solar thermal or powering electronics at mid-range temperatures (e.g., 500-600 K), fractionally filled materials, such as the Au$_{0.125}$(CoSb$_3$)$_4$ system, show the most promise for improved thermoelectric behavior.  

We note that Ag performs well only for the CoP$_3$ system, while Au performs well only for the CoAs$_3$ and CoSb$_3$ systems. Thus, it appears that filler size is important only when it exceeds a threshold percentage of the void radius. Since Au is the smallest filler considered in this study, it does not fulfill this criterion for any Co-based skutterudite.  La, being the largest filler considered in this study, does fulfill this criterion for all Co-based skutterudites and exhibits good performance in all cases. However, higher filler weights are important when considering systems with larger voids, and will likely have an even greater impact on thermal behavior.

The electronegativity of the filler may play an even greater role in determining the electrical behavior of the system. La, with an electronegativity of 1.1 on the Pauling scale, readily donates electrons to the lattice, which results in enhanced electrical properties. On the other hand, Au and Ag atoms are more selective in their willingness to donate electrons. Au has an electronegativity of 2.4 while Ag has an electronegativity of 1.9 on the Pauling scale. The electronegativity of Ag is much closer to that of the M atoms in the skutterudite (1.9-2.1 on the Pauling scale), while the difference in electronegativity between Au and the M atoms is much larger. In fact, since the electronegativity of Ag is so close to that of the M atoms, it is a poor filler in all but the phosphide skutterudites. On the other hand, Au is a good filler in arsenide and antimonide skutterudites, where the electronegativity difference is 0.4 and 0.5, respectively, but poor in the phosphides. Thus, we propose that the absolute difference in electronegativity between the filler and host atoms, combined with the atomic weight of the filler atom, are important considerations for designing improved thermoelectric materials. 

We now examine in greater depth four systems that exhibit improved electrical transport properties upon filling:

\begin{itemize}
\item 12.5\% Au filled CoSb$_3$
\item 100\% Au filled CoSb$_3$
\item 100\% Au filled CoAs$_3$
\item 100\% Ag filled CoP$_3$
\end{itemize}

Figure \ref{fig7_bestcases_bsdos} shows the band structures and density of states for these systems. Figures \ref{fig8_bestcases_au_properties}-\ref{fig9_bestcases_ag_properties} show the computed values of the Seebeck coefficient, $S$, and electrical conductivity, $\sigma$, as a function of temperature for these systems.  Two observations may be made:

\begin{itemize}
\item From the density of states, it seems that there is no appreciable band gap in the fully filled skutterudites, Au(CoSb$_3$)$_4$, Au(CoAs$_3$)$_4$, and Ag(CoP$_3$)$_4$, so it would appear likely that these compounds would be conducting and would not make good thermoelectric materials. However, in the case of Au(CoAs$_3$)$_4$ and Ag(CoP$_3$)$_4$, the band structure indicates that the filler creates an indirect band gap of -0.8 eV and -0.45 eV, respectively; these indirect band gaps are seen between the $P$ and $\Gamma$ $k$-points. We expect that these materials would therefore exhibit semimetallic rather than metallic properties. The sign of the $S$ values indicate that Au(CoAs$_3$)$_4$ exhibits p-type behavior while Ag(CoP$_3$)$_4$ exhibits n-type behavior.  These materials should thus be further studied to ascertain their thermal conductivities.
\item Filling CoSb$_3$ with small fractions of Au does not appreciably alter the narrow band gap. In Au$_{0.125}$(CoSb$_3$)$_4$, $S$ is greatly enhanced while $\sigma$ experiences only a slight increase. The resulting undoped, Au filled compound is a p-type semiconductor with high $S$ values.  To improve the values of $\sigma$, further analysis of doping was performed to determine whether enhancement of electrical transport properties is possible even at low Au filling fractions; this will be described in Section \ref{skut_symm}.
\end{itemize}

Figure \ref{fig_thermodynamics} shows a complete thermodynamics cycle for the formation of the minimally filled Au$_{0.125}$CoSb$_3$ system from CoSb$_3$.
\begin{itemize}
\item First, we calculated the total energy of unfilled CoSb$_3$ (-646.6 eV), and set this to be our reference point.
\item Second, we calculated the total energy of 12.5\% Au filled CoSb$_3$ (-648.9 eV).
\item Third, we calculated the total energy of unfilled CoSb$_3$ with the same void radius as 12.5\% Au  filled CoSb$_3$, by simply removing the Au filler atom and performing an energy minimization without geometry optimization.
\end{itemize}
By subtracting the total energy of the Au filler atom (-0.1 eV) from the system, we see that the formation of the filled compound is thermodynamically favorable, with an overall energy of formation of -2.2 eV.  Thus, it should be theoretically possible to synthesize partially Au filled skutterudites and experimentally characterize its properties for use in thermoelectric devices.

We also performed an analysis of the maximum filling fraction possible based on previously established methods \cite{Shi_FFL_2005}. This analysis tells us that, depending on the structure of the alternate stable phase AuSb$_2$, a maximum filling fraction of anywhere between 6\% (Fm3m symmetry) to 100\% (P2$_1\bar{c}$ symmetry) is possible without doping. Further information about secondary phases is necessary to more accurately determine the maximum filling fraction.

\subsection{\label{skut_symm2}Analysis of unit cell symmetry}

We previously presented our calculated electrical transport properties for the filled skutterudite systems exhibiting Im$\bar{3}$ symmetry in their unit cells, so we had omitted the 25\%, 50\%, and 75\% filling fractions in Figure \ref{fig6_powerfactors_all}. For Au$_x$(CoSb$_3$)$_4$, the 25\% filling fraction has three possible symmetry configurations: R$\bar{3}$, Cmmm, and Pmmm, the 50\% filling fraction has five possible symmetries, and the 75\% filling fraction has three possible symmetries. Since the 12.5\% filling fraction had been shown in Section \ref{skut_symm} to exhibit improved electrical transport properties, we were especially interested in exploring all of the possible symmetries for the 25\% filled structure, but in the case of the 50\% and 75\% filled structures, we chose instead to perform calculations for only one symmetry group each, which were Pm$\bar{3}$ and Cmmm, respectively.  We note that Im$\bar{3}$ is considered a high symmetry group, so the other symmetry groups being considered are lower in symmetry than Im$\bar{3}$ (Figure \ref{fig0_syms}).  In Figure \ref{fig12_au_all}, we plotted the power factor, $S^2 \sigma$, of CoSb$_3$ as a function of Au filling fraction at different unit cell symmetries corresponding to the filling fraction.  We observe that visible distortions in the overall trends are indeed observed at the 25\%, 50\%, and 75\% Au filling fractions, compared to those computed with Im$\bar{3}$ symmetry.  

To quantify the effect of symmetry on electrical transport properties, the 25\% Au filled system was re-optimized at all of its possible symmetries -- Cmmm (orthorhombic dipyramidal), Pmmm (orthorhombic dipyramidal), and R$\bar{3}$ (trigonal rhombohedral). From Figure \ref{fig12_au_all}, we see that the R$\bar{3}$ symmetry for the unit cell results in the largest power factors.  We conclude that the pattern in which the voids are filled is important for achieving favorable electrical transport properties, but stress that the full range of predicted values is likely to be observed experimentally when symmetry is difficult to control.  Figure \ref{fig12_au_all} also suggests that 25\% Au filling at Cmmm symmetry may exhibit a higher power factor than 12.5\% Au filling at Im$\bar{3}$ symmetry, and this prediction is confirmed by the results shown in Figure \ref{fig13_au_25} for power factor as a function of temperature.

Figure \ref{fig14_au_fermisweep} shows the $S$, $\sigma$, and $S^2\sigma$ values for the 12.5\%, 25\%, and 100\% Au filled CoSb$_3$ systems at 500 K. The graphs show values for doping levels from -0.5 eV to +0.5 eV around the Fermi energy, as simulated using the rigid band model. From these graphs we note that filling with Au creates a second hump in the $S$ value which shifts further away from the fermi energy with increasing filler concentration. This second hump also results in the highest values of $S^2\sigma$. The greatest potential for thermoelectric behavior, in terms of the $\text{tr}  \left( S \right)$ value, is observed in the 25\% filled compound, while directionally, it is observed for the $S_{xx}$ component in the 100\% filled compound. Fully filling the lattice with Au appears to unfairly dope the host material by a suboptimal amount (i.e., too little or too much) into the valence band, so that the power factor is compromised at 500 K.  However, further improvement in the power factor may be achieved by doping the material.  We conclude that low fractional filling with Au appears to induce promising p-type behavior and provides a good complement to n-type Lanthanum filled materials in p-n junction devices. 

\section{\label{sec_conc}Conclusions}
\begin{figure}
\includegraphics[width=150mm]{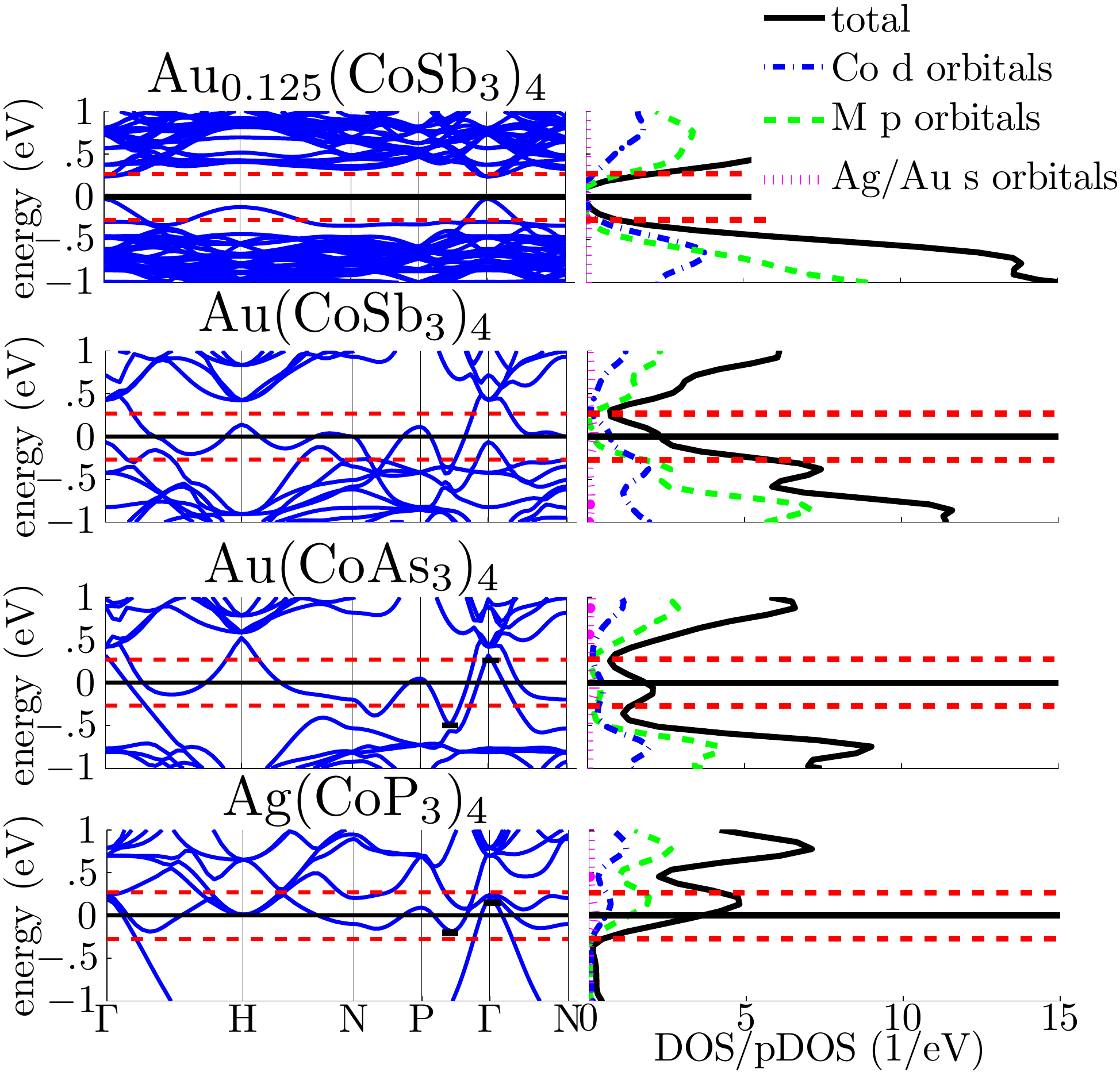}
\caption{Band structures and partial density of states for best-performing Ag and Au filled Co-based skutterudites: 12.5\% and 100\% Au filled CoSb$_3$, 100\% Au filled CoAs$_3$, and 100\% Ag filled CoP$_3$.\label{fig7_bestcases_bsdos}}
\end{figure}

\begin{figure}
\includegraphics[width=130mm]{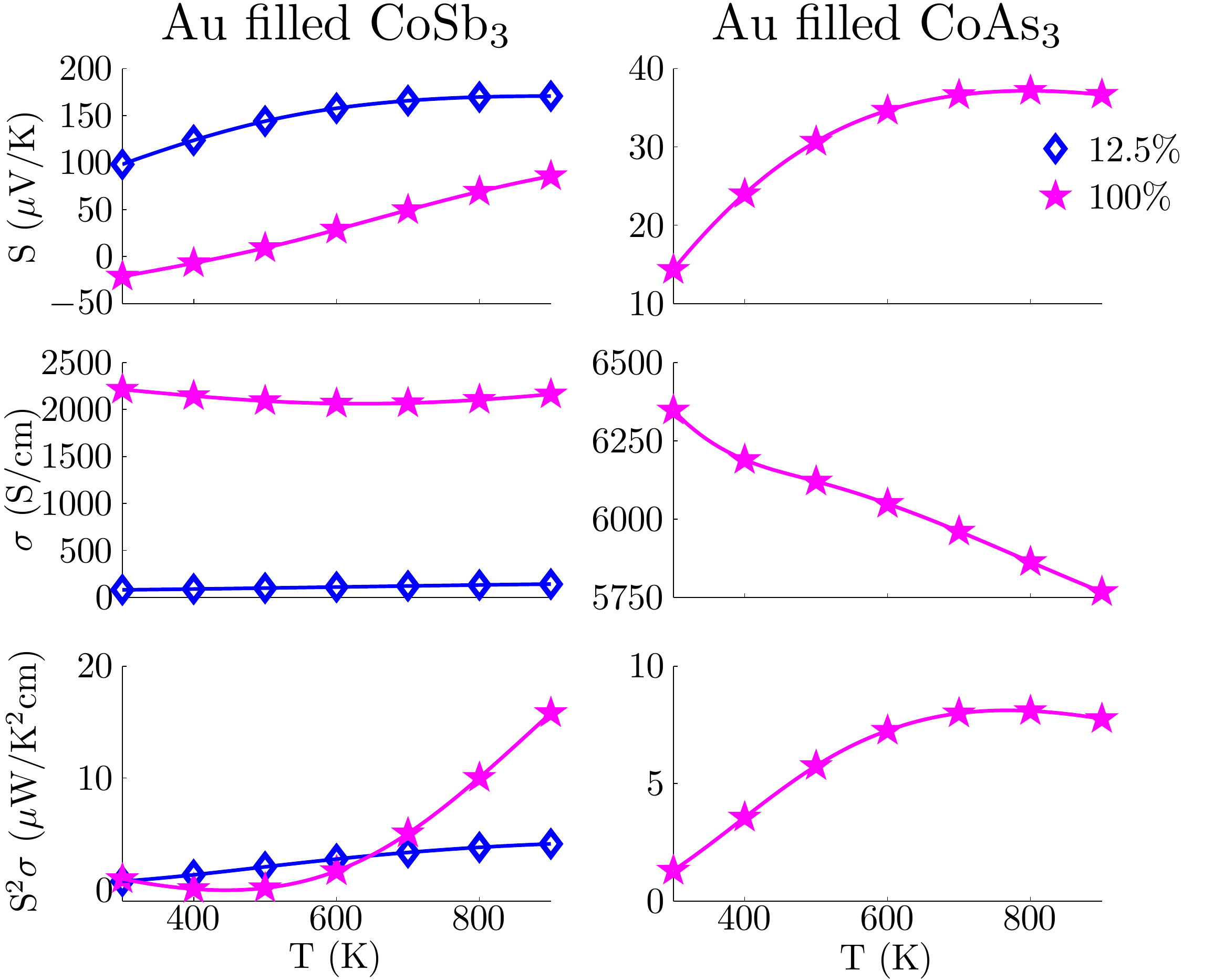}
\caption{Electrical transport properties, $S$, $\sigma$, and $S^2 \sigma$, as a function of temperature for 12.5\% and 100\% Au filled CoSb$_3$ and 100\% Au filled CoAs$_3$.\label{fig8_bestcases_au_properties}}
\end{figure}

\begin{figure}
\includegraphics[width=65mm]{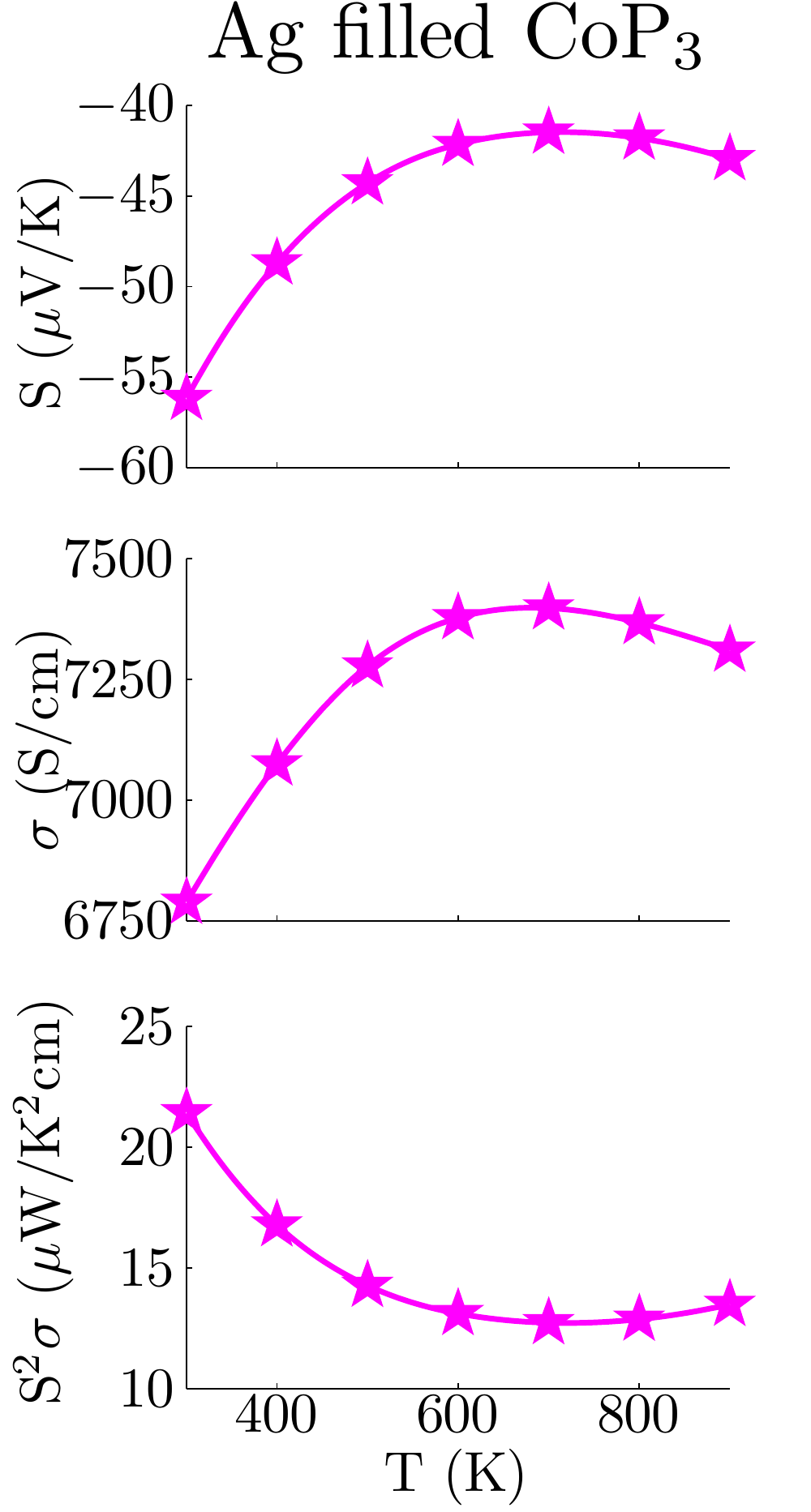}
\caption{Electrical transport properties, $S$, $\sigma$, and $S^2 \sigma$, as a function of temperature for 100\% Ag filled CoP$_3$.\label{fig9_bestcases_ag_properties}}
\end{figure}

\begin{figure}
\includegraphics[width=110mm]{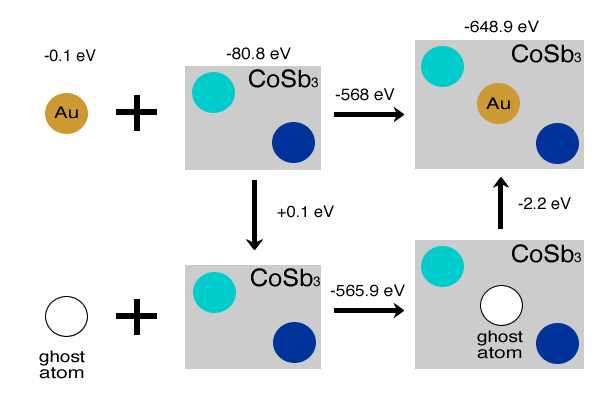}
\caption{Thermodynamic cycle indicating the thermodynamically downhill energy of formation for Au filled CoSb$_3$.\label{fig_thermodynamics}}
\end{figure}

\begin{figure}
\includegraphics[height=130mm]{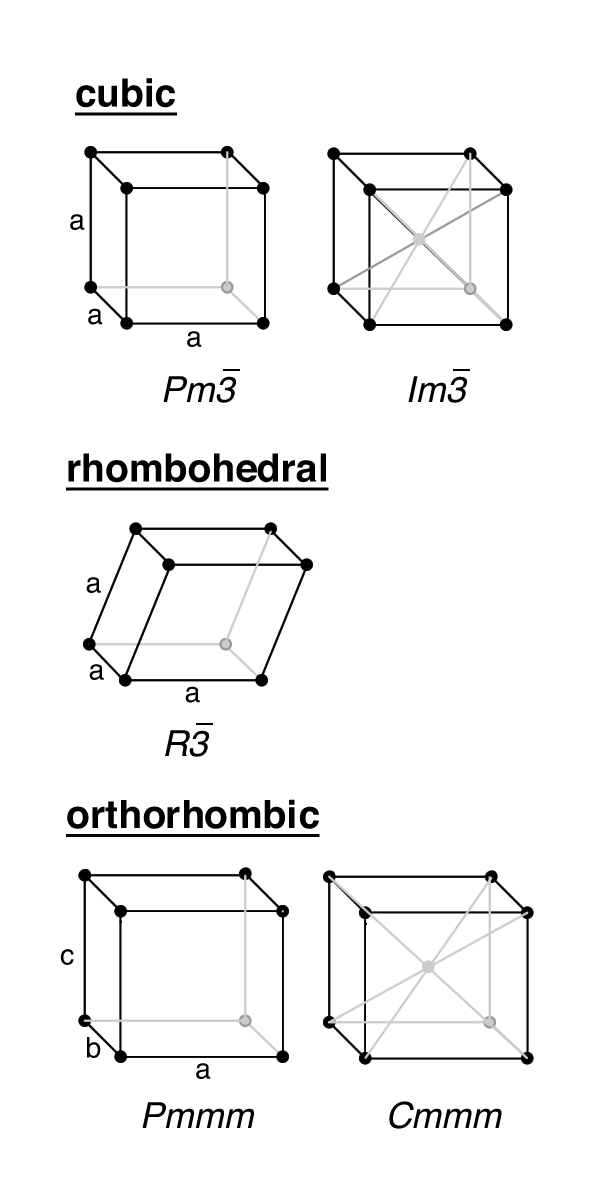}
\caption{Symmetries of filled skutterudite systems. Cubic systems have the highest symmetry, followed by rhombohedral and then orthorhombic systems.\label{fig0_syms}}
\end{figure}

\begin{figure}
\includegraphics[width=130mm]{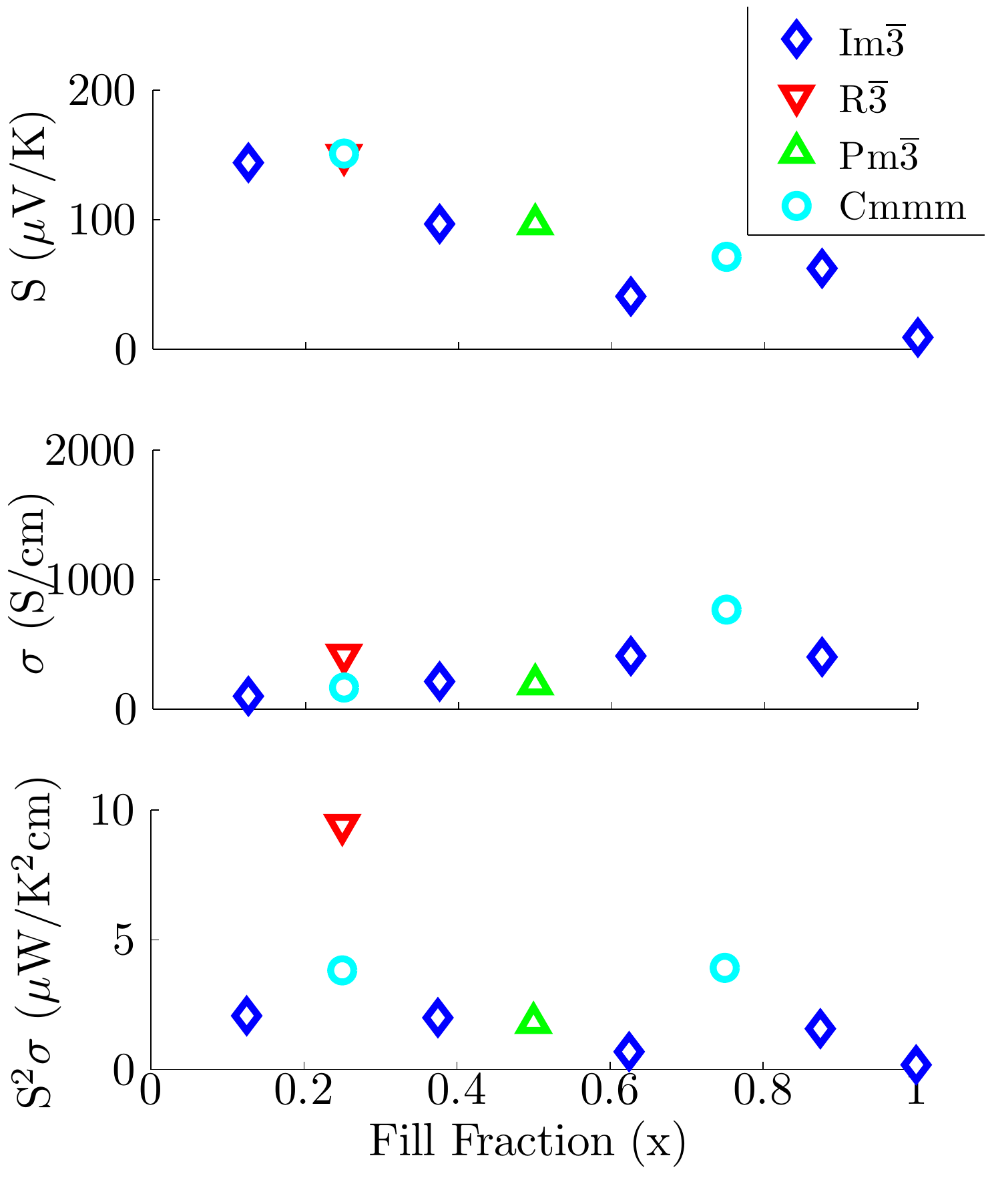}
\caption{Electrical transport properties, $S$, $\sigma$, $S^2 \sigma$, as a function of Au filling fraction, $x$, for all symmetries computed for Au$_x$(CoSb$_3$)$_4$.\label{fig12_au_all}}
\end{figure}

\begin{figure}
\includegraphics[width=130mm]{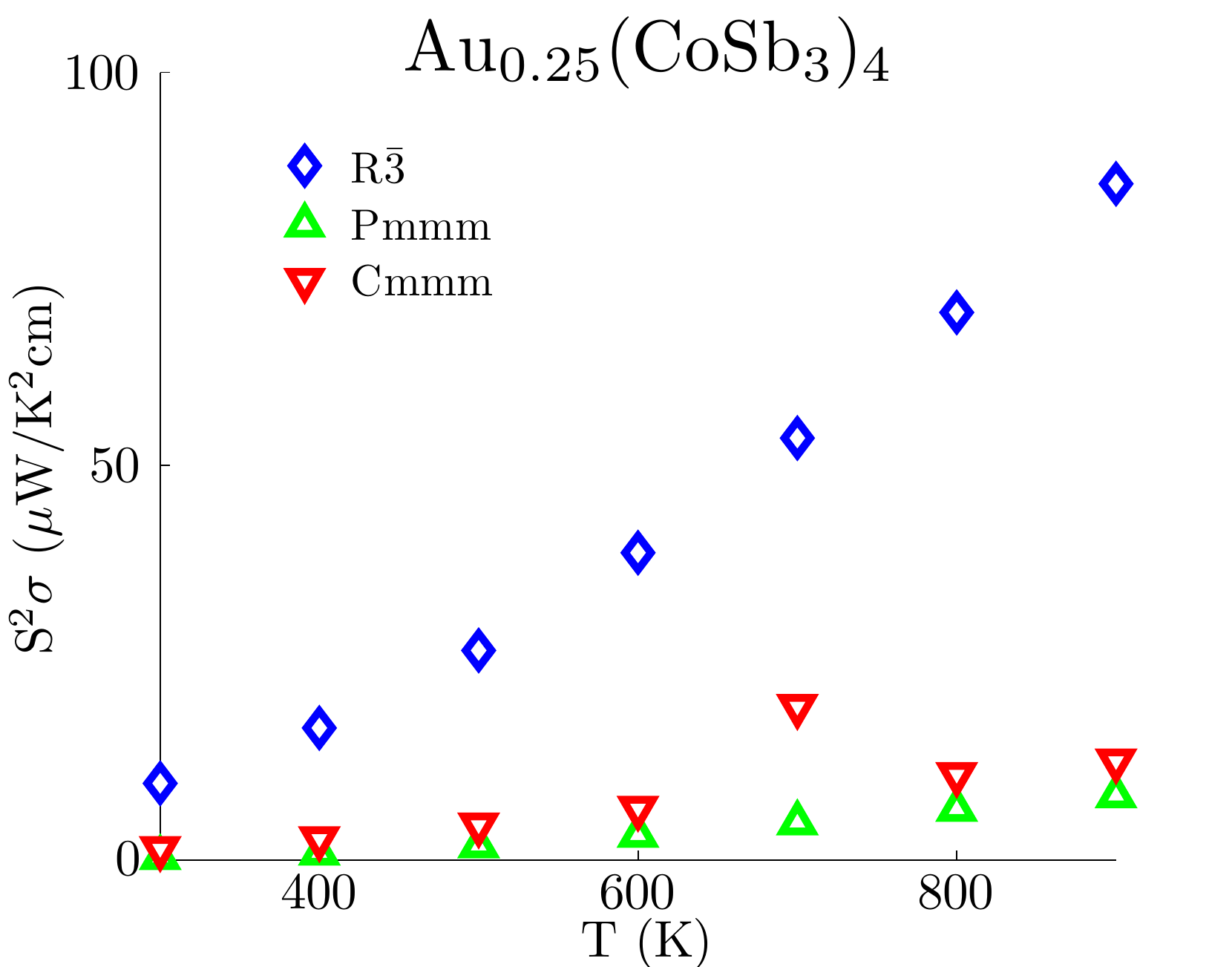}
\caption{Thermoelectric power factor, $S^2 \sigma$, as a function of temperature for 25\% Au filled CoSb$_3$.\label{fig13_au_25}}
\end{figure}

\begin{figure}
\includegraphics[width=150mm]{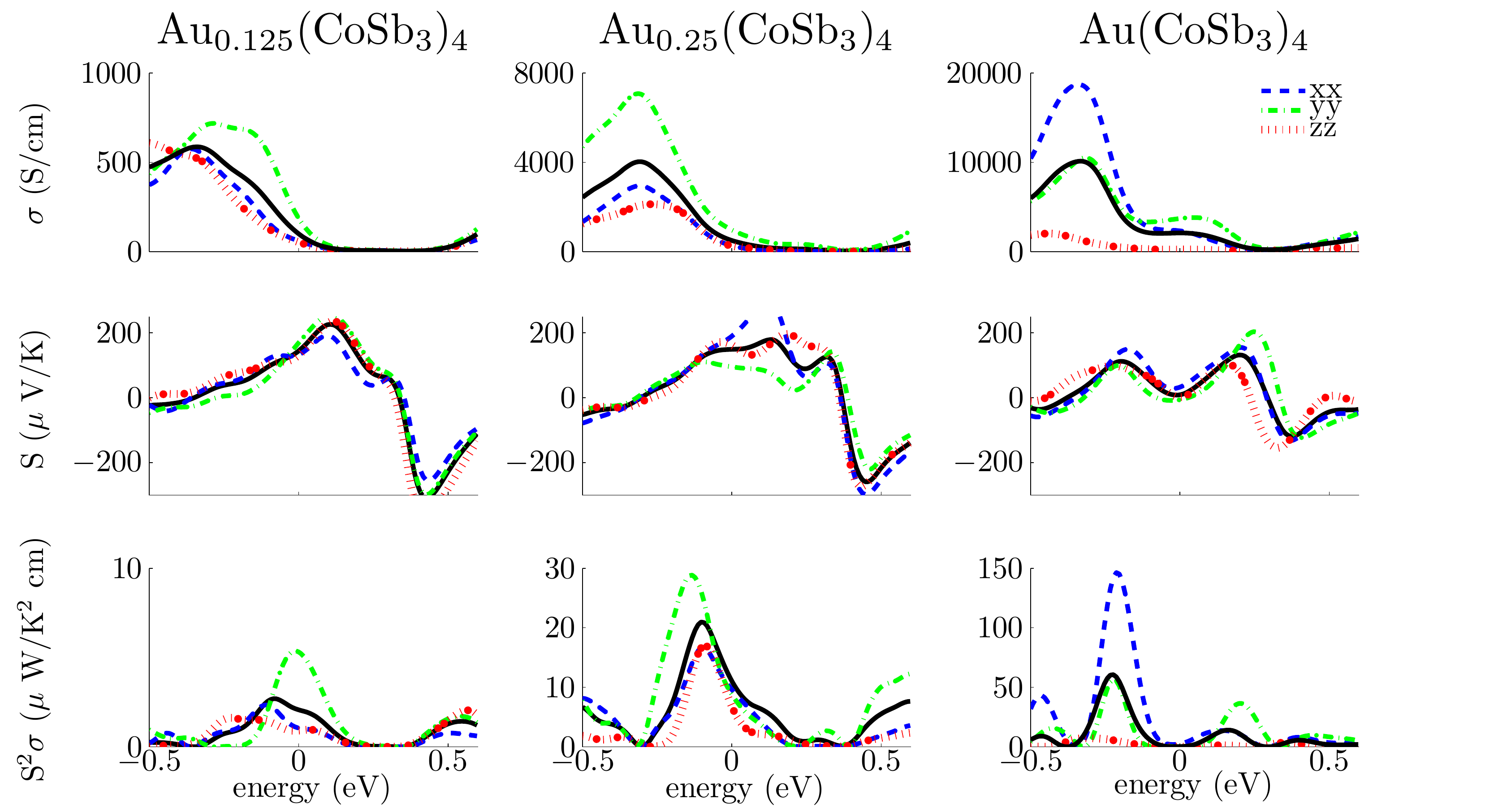}
\caption{Directional (i.e., $xx$, $yy$, and $zz$) electrical transport properties, $S$, $\sigma$, and $S^2 \sigma$, as a function of energy for 12.5\% (Im$\bar{3}$), 25\% (R$\bar{3}$), and 100\% (Im$\bar{3}$) Au filled CoSb$_3$.\label{fig14_au_fermisweep}}
\end{figure}

We have shown that density functional theory calculations following the Projector-Augmented Wave approach, in conjunction with the momentum matrix method for calculating electrical transport properties, may be used successfully to obtain predictive trends in the performance of Co-based skutterudites over a temperature range of 300-900 K. Au filling results in p-type behavior whereas Ag filling results in n-type behavior. When considering the properties of candidate fillers, the relative weight and electronegativity appear to affect electrical transport properties more than the absolute size of the filler atom. These properties are evident when considering that the hybridization of the filler atom's $s$ electrons is most emphasized in the fully filled skutterudites, as demonstrated in the density of states around the Fermi level in Figure \ref{fig7_bestcases_bsdos}. Therefore, in these fully filled systems, a resonance quality arises in the lattice where the coordination of electrons results in greatly enhanced electrical conductivities; this enhancement is most notable at high temperatures.  Based on our calculations, fully Au filled CoSb$_3$ would make an excellent thermoelectric material at temperatures above 600 K, but it may be difficult to synthesize this material in a cost-effective manner \cite{PhysRevB.74.153202,PhysRevB.75.235208} and it may even exhibit undesirable (semi)metallic behavior, as suggested by our band structure calculations. Thus, we propose that low Au filling fractions (e.g., 25\% or 12.5\%) offer the optimal balance between improved performance and economical value. In terms of building an efficient thermoelectric device with a p-n junction, we suggest that coupling p-type Au filled CoSb$_3$ with n-type Lanthanum (or other rare earth element) filled CoSb$_3$ would favorably result in a device with optimal performance yet minimal lattice strain at the heterojunction.

\begin{acknowledgments}
This research was supported in part by the National Science Foundation through TeraGrid resources provided by the Texas Advanced Computing Cluster under grant number TG-CTS110011, and by the Mr. and Mrs. Spencer T. Olin Fellowship program at Washington University through a graduate fellowship provided to Maria Stoica.
\end{acknowledgments}

\clearpage


\begin{thebibliography}{36}%
\makeatletter
\providecommand \@ifxundefined [1]{%
 \@ifx{#1\undefined}
}%
\providecommand \@ifnum [1]{%
 \ifnum #1\expandafter \@firstoftwo
 \else \expandafter \@secondoftwo
 \fi
}%
\providecommand \@ifx [1]{%
 \ifx #1\expandafter \@firstoftwo
 \else \expandafter \@secondoftwo
 \fi
}%
\providecommand \natexlab [1]{#1}%
\providecommand \enquote  [1]{``#1''}%
\providecommand \bibnamefont  [1]{#1}%
\providecommand \bibfnamefont [1]{#1}%
\providecommand \citenamefont [1]{#1}%
\providecommand \href@noop [0]{\@secondoftwo}%
\providecommand \href [0]{\begingroup \@sanitize@url \@href}%
\providecommand \@href[1]{\@@startlink{#1}\@@href}%
\providecommand \@@href[1]{\endgroup#1\@@endlink}%
\providecommand \@sanitize@url [0]{\catcode `\\12\catcode `\$12\catcode
  `\&12\catcode `\#12\catcode `\^12\catcode `\_12\catcode `\%12\relax}%
\providecommand \@@startlink[1]{}%
\providecommand \@@endlink[0]{}%
\providecommand \url  [0]{\begingroup\@sanitize@url \@url }%
\providecommand \@url [1]{\endgroup\@href {#1}{\urlprefix }}%
\providecommand \urlprefix  [0]{URL }%
\providecommand \Eprint [0]{\href }%
\providecommand \doibase [0]{http://dx.doi.org/}%
\providecommand \selectlanguage [0]{\@gobble}%
\providecommand \bibinfo  [0]{\@secondoftwo}%
\providecommand \bibfield  [0]{\@secondoftwo}%
\providecommand \translation [1]{[#1]}%
\providecommand \BibitemOpen [0]{}%
\providecommand \bibitemStop [0]{}%
\providecommand \bibitemNoStop [0]{.\EOS\space}%
\providecommand \EOS [0]{\spacefactor3000\relax}%
\providecommand \BibitemShut  [1]{\csname bibitem#1\endcsname}%
\let\auto@bib@innerbib\@empty
\bibitem [{\citenamefont {Bell}(2008)}]{Bell_Science_ThermoelectricApps}%
  \BibitemOpen
  \bibfield  {author} {\bibinfo {author} {\bibfnamefont {L.~E.}\ \bibnamefont
  {Bell}},\ }\href {\doibase 10.1126/science.1158899} {\bibfield  {journal}
  {\bibinfo  {journal} {Science}\ }\textbf {\bibinfo {volume} {321}},\ \bibinfo
  {pages} {1457} (\bibinfo {year} {2008})}\BibitemShut {NoStop}%
\bibitem [{\citenamefont {Yang}(2005)}]{Yang_IEEEroc_ThermoelectricsAppsAuto}%
  \BibitemOpen
  \bibfield  {author} {\bibinfo {author} {\bibfnamefont {J.}~\bibnamefont
  {Yang}},\ }in\ \href {\doibase 10.1109/ICT.2005.1519911} {\emph {\bibinfo
  {booktitle} {ICT 2005. 24th International Conference on Thermoelectrics}}}\
  (\bibinfo {year} {2005})\ pp.\ \bibinfo {pages} {170--174}\BibitemShut
  {NoStop}%
\bibitem [{\citenamefont {Tritt}\ \emph {et~al.}(2008)\citenamefont {Tritt},
  \citenamefont {B{\"o}ttner},\ and\ \citenamefont
  {Chen}}]{Tritt_MRSBul_ThermoelectricSolar}%
  \BibitemOpen
  \bibfield  {author} {\bibinfo {author} {\bibfnamefont {T.}~\bibnamefont
  {Tritt}}, \bibinfo {author} {\bibfnamefont {H.}~\bibnamefont {B{\"o}ttner}},
  \ and\ \bibinfo {author} {\bibfnamefont {L.}~\bibnamefont {Chen}},\ }\href
  {\doibase 10.1557/mrs2008.73} {\bibfield  {journal} {\bibinfo  {journal} {MRS
  Bulletin}\ }\textbf {\bibinfo {volume} {33}},\ \bibinfo {pages} {366}
  (\bibinfo {year} {2008})}\BibitemShut {NoStop}%
\bibitem [{\citenamefont {Tritt}\ and\ \citenamefont
  {Subramanian}(2006)}]{Tritt_MRSBul_ThermoelectricPhenomenaApps}%
  \BibitemOpen
  \bibfield  {author} {\bibinfo {author} {\bibfnamefont {T.~M.}\ \bibnamefont
  {Tritt}}\ and\ \bibinfo {author} {\bibfnamefont {M.~A.}\ \bibnamefont
  {Subramanian}},\ }\href {\doibase 10.1557/mrs2006.44} {\bibfield  {journal}
  {\bibinfo  {journal} {MRS Bulletin}\ }\textbf {\bibinfo {volume} {31}},\
  \bibinfo {pages} {188} (\bibinfo {year} {2006})}\BibitemShut {NoStop}%
\bibitem [{\citenamefont {Mahan}\ and\ \citenamefont
  {Sofo}(1996)}]{Mahan23071996}%
  \BibitemOpen
  \bibfield  {author} {\bibinfo {author} {\bibfnamefont {G.~D.}\ \bibnamefont
  {Mahan}}\ and\ \bibinfo {author} {\bibfnamefont {J.~O.}\ \bibnamefont
  {Sofo}},\ }\href {http://www.pnas.org/content/93/15/7436.abstract} {\bibfield
   {journal} {\bibinfo  {journal} {Proceedings of the National Academy of
  Sciences}\ }\textbf {\bibinfo {volume} {93}},\ \bibinfo {pages} {7436}
  (\bibinfo {year} {1996})},\ \Eprint
  {http://arxiv.org/abs/http://www.pnas.org/content/93/15/7436.full.pdf+html}
  {http://www.pnas.org/content/93/15/7436.full.pdf+html} \BibitemShut {NoStop}%
\bibitem [{\citenamefont {Sales}\ \emph {et~al.}(1997)\citenamefont {Sales},
  \citenamefont {Mandrus}, \citenamefont {Chakoumakos}, \citenamefont
  {Keppens},\ and\ \citenamefont {Thompson}}]{Sales_PRB_LaCeFilledCoSb3}%
  \BibitemOpen
  \bibfield  {author} {\bibinfo {author} {\bibfnamefont {B.~C.}\ \bibnamefont
  {Sales}}, \bibinfo {author} {\bibfnamefont {D.}~\bibnamefont {Mandrus}},
  \bibinfo {author} {\bibfnamefont {B.~C.}\ \bibnamefont {Chakoumakos}},
  \bibinfo {author} {\bibfnamefont {V.}~\bibnamefont {Keppens}}, \ and\
  \bibinfo {author} {\bibfnamefont {J.~R.}\ \bibnamefont {Thompson}},\ }\href
  {\doibase 10.1103/PhysRevB.56.15081} {\bibfield  {journal} {\bibinfo
  {journal} {Physical Review B}\ }\textbf {\bibinfo {volume} {56}},\ \bibinfo
  {pages} {15081} (\bibinfo {year} {1997})}\BibitemShut {NoStop}%
\bibitem [{\citenamefont {Lamberton}\ \emph {et~al.}(2002)\citenamefont
  {Lamberton}, \citenamefont {Bhattacharya}, \citenamefont {Littleton},
  \citenamefont {Kaeser}, \citenamefont {Tedstrom}, \citenamefont {Tritt},
  \citenamefont {Yang},\ and\ \citenamefont
  {Nolas}}]{Lamberton_APL_EuFilledCoSb3}%
  \BibitemOpen
  \bibfield  {author} {\bibinfo {author} {\bibfnamefont {G.~A.}\ \bibnamefont
  {Lamberton}}, \bibinfo {author} {\bibfnamefont {S.}~\bibnamefont
  {Bhattacharya}}, \bibinfo {author} {\bibfnamefont {R.~T.}\ \bibnamefont
  {Littleton}}, \bibinfo {author} {\bibfnamefont {M.~A.}\ \bibnamefont
  {Kaeser}}, \bibinfo {author} {\bibfnamefont {R.~H.}\ \bibnamefont
  {Tedstrom}}, \bibinfo {author} {\bibfnamefont {T.~M.}\ \bibnamefont {Tritt}},
  \bibinfo {author} {\bibfnamefont {J.}~\bibnamefont {Yang}}, \ and\ \bibinfo
  {author} {\bibfnamefont {G.~S.}\ \bibnamefont {Nolas}},\ }\href
  {http://libproxy.wustl.edu/login?url=http://search.ebscohost.com/login.aspx?direct=true&db=aph&AN=5942532&site=ehost-live&scope=site}
  {\bibfield  {journal} {\bibinfo  {journal} {Applied Physics Letters}\
  }\textbf {\bibinfo {volume} {80}},\ \bibinfo {pages} {598} (\bibinfo {year}
  {2002})}\BibitemShut {NoStop}%
\bibitem [{\citenamefont {Nordstr{\"o}m}\ and\ \citenamefont
  {Singh}(1996)}]{PhysRevB.53.1103}%
  \BibitemOpen
  \bibfield  {author} {\bibinfo {author} {\bibfnamefont {L.}~\bibnamefont
  {Nordstr{\"o}m}}\ and\ \bibinfo {author} {\bibfnamefont {D.~J.}\ \bibnamefont
  {Singh}},\ }\href {\doibase 10.1103/PhysRevB.53.1103} {\bibfield  {journal}
  {\bibinfo  {journal} {Physical Review B}\ }\textbf {\bibinfo {volume} {53}},\
  \bibinfo {pages} {1103} (\bibinfo {year} {1996})}\BibitemShut {NoStop}%
\bibitem [{\citenamefont {Nolas}\ \emph {et~al.}(2006)\citenamefont {Nolas},
  \citenamefont {Poon},\ and\ \citenamefont
  {Kanatzidis}}]{Nolas_MRSBul_RecentDevelThermoelectrics}%
  \BibitemOpen
  \bibfield  {author} {\bibinfo {author} {\bibfnamefont {G.~S.}\ \bibnamefont
  {Nolas}}, \bibinfo {author} {\bibfnamefont {J.}~\bibnamefont {Poon}}, \ and\
  \bibinfo {author} {\bibfnamefont {M.}~\bibnamefont {Kanatzidis}},\ }\href
  {\doibase 10.1557/mrs2006.45} {\bibfield  {journal} {\bibinfo  {journal} {MRS
  Bulletin}\ }\textbf {\bibinfo {volume} {31}},\ \bibinfo {pages} {199}
  (\bibinfo {year} {2006})}\BibitemShut {NoStop}%
\bibitem [{\citenamefont {Chen}\ \emph {et~al.}(2001)\citenamefont {Chen},
  \citenamefont {Kawahara}, \citenamefont {Tang}, \citenamefont {Goto},
  \citenamefont {Hirai}, \citenamefont {Dyck}, \citenamefont {Chen},\ and\
  \citenamefont {Uher}}]{Chen_JAP_BaFilledCoSb3}%
  \BibitemOpen
  \bibfield  {author} {\bibinfo {author} {\bibfnamefont {L.~D.}\ \bibnamefont
  {Chen}}, \bibinfo {author} {\bibfnamefont {T.}~\bibnamefont {Kawahara}},
  \bibinfo {author} {\bibfnamefont {X.~F.}\ \bibnamefont {Tang}}, \bibinfo
  {author} {\bibfnamefont {T.}~\bibnamefont {Goto}}, \bibinfo {author}
  {\bibfnamefont {T.}~\bibnamefont {Hirai}}, \bibinfo {author} {\bibfnamefont
  {J.~S.}\ \bibnamefont {Dyck}}, \bibinfo {author} {\bibfnamefont
  {W.}~\bibnamefont {Chen}}, \ and\ \bibinfo {author} {\bibfnamefont
  {C.}~\bibnamefont {Uher}},\ }\href {\doibase 10.1063/1.1388162} {\bibfield
  {journal} {\bibinfo  {journal} {Journal of Applied Physics}\ }\textbf
  {\bibinfo {volume} {90}},\ \bibinfo {pages} {1864} (\bibinfo {year}
  {2001})}\BibitemShut {NoStop}%
\bibitem [{\citenamefont {Pei}\ \emph {et~al.}(2009)\citenamefont {Pei},
  \citenamefont {Yang}, \citenamefont {Chen}, \citenamefont {Zhang},
  \citenamefont {Salvador},\ and\ \citenamefont
  {Yang}}]{Pei_APL_NaFilledCoSb3}%
  \BibitemOpen
  \bibfield  {author} {\bibinfo {author} {\bibfnamefont {Y.~Z.}\ \bibnamefont
  {Pei}}, \bibinfo {author} {\bibfnamefont {J.}~\bibnamefont {Yang}}, \bibinfo
  {author} {\bibfnamefont {L.~D.}\ \bibnamefont {Chen}}, \bibinfo {author}
  {\bibfnamefont {W.}~\bibnamefont {Zhang}}, \bibinfo {author} {\bibfnamefont
  {J.~R.}\ \bibnamefont {Salvador}}, \ and\ \bibinfo {author} {\bibfnamefont
  {J.}~\bibnamefont {Yang}},\ }\href {\doibase 10.1063/1.3182800} {\bibfield
  {journal} {\bibinfo  {journal} {Applied Physics Letters}\ }\textbf {\bibinfo
  {volume} {95}},\ \bibinfo {eid} {042101} (\bibinfo {year}
  {2009})}\BibitemShut {NoStop}%
\bibitem [{\citenamefont {He}\ \emph {et~al.}(2006)\citenamefont {He},
  \citenamefont {Chen}, \citenamefont {Rosenfeld},\ and\ \citenamefont
  {Subramanian}}]{He_ChemMat_InFilledCoSb3}%
  \BibitemOpen
  \bibfield  {author} {\bibinfo {author} {\bibfnamefont {T.}~\bibnamefont
  {He}}, \bibinfo {author} {\bibfnamefont {J.}~\bibnamefont {Chen}}, \bibinfo
  {author} {\bibfnamefont {H.~D.}\ \bibnamefont {Rosenfeld}}, \ and\ \bibinfo
  {author} {\bibfnamefont {M.~A.}\ \bibnamefont {Subramanian}},\ }\href
  {\doibase 10.1021/cm052055b} {\bibfield  {journal} {\bibinfo  {journal}
  {Chemistry of Materials}\ }\textbf {\bibinfo {volume} {18}},\ \bibinfo
  {pages} {759} (\bibinfo {year} {2006})}\BibitemShut {NoStop}%
\bibitem [{\citenamefont {Yang}\ \emph {et~al.}(2007)\citenamefont {Yang},
  \citenamefont {Zhang}, \citenamefont {Bai}, \citenamefont {Mei},\ and\
  \citenamefont {Chen}}]{Yang_APL_DoubleFilledCoSb3}%
  \BibitemOpen
  \bibfield  {author} {\bibinfo {author} {\bibfnamefont {J.}~\bibnamefont
  {Yang}}, \bibinfo {author} {\bibfnamefont {W.}~\bibnamefont {Zhang}},
  \bibinfo {author} {\bibfnamefont {S.~Q.}\ \bibnamefont {Bai}}, \bibinfo
  {author} {\bibfnamefont {Z.}~\bibnamefont {Mei}}, \ and\ \bibinfo {author}
  {\bibfnamefont {L.~D.}\ \bibnamefont {Chen}},\ }\href {\doibase
  10.1063/1.2737422} {\bibfield  {journal} {\bibinfo  {journal} {Applied
  Physics Letters}\ }\textbf {\bibinfo {volume} {90}},\ \bibinfo {pages}
  {192111 } (\bibinfo {year} {2007})}\BibitemShut {NoStop}%
\bibitem [{\citenamefont {Madsen}\ and\ \citenamefont
  {Singh}(2006)}]{Madsen_CPC_Boltztrapp}%
  \BibitemOpen
  \bibfield  {author} {\bibinfo {author} {\bibfnamefont {G.~K.}\ \bibnamefont
  {Madsen}}\ and\ \bibinfo {author} {\bibfnamefont {D.~J.}\ \bibnamefont
  {Singh}},\ }\href {\doibase 10.1016/j.cpc.2006.03.007} {\bibfield  {journal}
  {\bibinfo  {journal} {Computer Physics Communications}\ }\textbf {\bibinfo
  {volume} {175}},\ \bibinfo {pages} {67 } (\bibinfo {year}
  {2006})}\BibitemShut {NoStop}%
\bibitem [{\citenamefont {Scheidemantel}\ \emph {et~al.}(2003)\citenamefont
  {Scheidemantel}, \citenamefont {Ambrosch-Draxl}, \citenamefont {Thonhauser},
  \citenamefont {Badding},\ and\ \citenamefont
  {Sofo}}]{Scheidemantel_PRB_TransportCoeffsFirstPrinciples}%
  \BibitemOpen
  \bibfield  {author} {\bibinfo {author} {\bibfnamefont {T.~J.}\ \bibnamefont
  {Scheidemantel}}, \bibinfo {author} {\bibfnamefont {C.}~\bibnamefont
  {Ambrosch-Draxl}}, \bibinfo {author} {\bibfnamefont {T.}~\bibnamefont
  {Thonhauser}}, \bibinfo {author} {\bibfnamefont {J.~V.}\ \bibnamefont
  {Badding}}, \ and\ \bibinfo {author} {\bibfnamefont {J.~O.}\ \bibnamefont
  {Sofo}},\ }\href {\doibase 10.1103/PhysRevB.68.125210} {\bibfield  {journal}
  {\bibinfo  {journal} {Physical Review B}\ }\textbf {\bibinfo {volume} {68}},\
  \bibinfo {pages} {125210} (\bibinfo {year} {2003})}\BibitemShut {NoStop}%
\bibitem [{\citenamefont {Adolph}\ \emph {et~al.}(2001)\citenamefont {Adolph},
  \citenamefont {Furthm\"uller},\ and\ \citenamefont
  {Bechstedt}}]{Furthmuller_PRB_OpticalPropertiesUsingPAW}%
  \BibitemOpen
  \bibfield  {author} {\bibinfo {author} {\bibfnamefont {B.}~\bibnamefont
  {Adolph}}, \bibinfo {author} {\bibfnamefont {J.}~\bibnamefont
  {Furthm\"uller}}, \ and\ \bibinfo {author} {\bibfnamefont {F.}~\bibnamefont
  {Bechstedt}},\ }\href {\doibase 10.1103/PhysRevB.63.125108} {\bibfield
  {journal} {\bibinfo  {journal} {Physical Review B}\ }\textbf {\bibinfo
  {volume} {63}},\ \bibinfo {pages} {125108} (\bibinfo {year}
  {2001})}\BibitemShut {NoStop}%
\bibitem [{\citenamefont {Kageshima}\ and\ \citenamefont
  {Shiraishi}(1997)}]{Kageshima_PRB_MomMatrixCalcPP}%
  \BibitemOpen
  \bibfield  {author} {\bibinfo {author} {\bibfnamefont {H.}~\bibnamefont
  {Kageshima}}\ and\ \bibinfo {author} {\bibfnamefont {K.}~\bibnamefont
  {Shiraishi}},\ }\href {\doibase 10.1103/PhysRevB.56.14985} {\bibfield
  {journal} {\bibinfo  {journal} {Physical Review B}\ }\textbf {\bibinfo
  {volume} {56}},\ \bibinfo {pages} {14985} (\bibinfo {year}
  {1997})}\BibitemShut {NoStop}%
\bibitem [{\citenamefont {Kresse}\ and\ \citenamefont
  {Hafner}(1993)}]{Kresse_CompMatSci_VASPRef0}%
  \BibitemOpen
  \bibfield  {author} {\bibinfo {author} {\bibfnamefont {G.}~\bibnamefont
  {Kresse}}\ and\ \bibinfo {author} {\bibfnamefont {J.}~\bibnamefont
  {Hafner}},\ }\href@noop {} {\bibfield  {journal} {\bibinfo  {journal}
  {Physical Review B}\ }\textbf {\bibinfo {volume} {47}},\ \bibinfo {pages}
  {558} (\bibinfo {year} {1993})}\BibitemShut {NoStop}%
\bibitem [{\citenamefont {Kresse}\ and\ \citenamefont
  {Hafner}(1994)}]{Kresse_PRB_VASPRef1}%
  \BibitemOpen
  \bibfield  {author} {\bibinfo {author} {\bibfnamefont {G.}~\bibnamefont
  {Kresse}}\ and\ \bibinfo {author} {\bibfnamefont {J.}~\bibnamefont
  {Hafner}},\ }\href@noop {} {\bibfield  {journal} {\bibinfo  {journal}
  {Physical Review B}\ }\textbf {\bibinfo {volume} {49}},\ \bibinfo {pages}
  {14251} (\bibinfo {year} {1994})}\BibitemShut {NoStop}%
\bibitem [{\citenamefont {Kresse}\ and\ \citenamefont
  {Furthm{\"u}ller}(1996{\natexlab{a}})}]{Kresse_PRB_VASPRef3}%
  \BibitemOpen
  \bibfield  {author} {\bibinfo {author} {\bibfnamefont {G.}~\bibnamefont
  {Kresse}}\ and\ \bibinfo {author} {\bibfnamefont {J.}~\bibnamefont
  {Furthm{\"u}ller}},\ }\href@noop {} {\bibfield  {journal} {\bibinfo
  {journal} {Physical Review B}\ }\textbf {\bibinfo {volume} {54}},\ \bibinfo
  {pages} {11169} (\bibinfo {year} {1996}{\natexlab{a}})}\BibitemShut {NoStop}%
\bibitem [{\citenamefont {Kresse}\ and\ \citenamefont
  {Furthm{\"u}ller}(1996{\natexlab{b}})}]{Kresse_CompMatSci_VASPRef2}%
  \BibitemOpen
  \bibfield  {author} {\bibinfo {author} {\bibfnamefont {G.}~\bibnamefont
  {Kresse}}\ and\ \bibinfo {author} {\bibfnamefont {J.}~\bibnamefont
  {Furthm{\"u}ller}},\ }\href@noop {} {\bibfield  {journal} {\bibinfo
  {journal} {Computational Materials Science}\ }\textbf {\bibinfo {volume}
  {6}},\ \bibinfo {pages} {15} (\bibinfo {year}
  {1996}{\natexlab{b}})}\BibitemShut {NoStop}%
\bibitem [{\citenamefont {Bl{\"o}chl}(1994)}]{Bloechl_PRB_PAWMethod}%
  \BibitemOpen
  \bibfield  {author} {\bibinfo {author} {\bibfnamefont {P.~E.}\ \bibnamefont
  {Bl{\"o}chl}},\ }\href@noop {} {\bibfield  {journal} {\bibinfo  {journal}
  {Physical Review B}\ }\textbf {\bibinfo {volume} {50}},\ \bibinfo {pages}
  {17953} (\bibinfo {year} {1994})}\BibitemShut {NoStop}%
\bibitem [{\citenamefont {Kresse}\ and\ \citenamefont
  {Joubert}(1999)}]{Kresse_PRB_PSSToPAW}%
  \BibitemOpen
  \bibfield  {author} {\bibinfo {author} {\bibfnamefont {G.}~\bibnamefont
  {Kresse}}\ and\ \bibinfo {author} {\bibfnamefont {D.}~\bibnamefont
  {Joubert}},\ }\href@noop {} {\bibfield  {journal} {\bibinfo  {journal}
  {Physical Review B}\ }\textbf {\bibinfo {volume} {59}},\ \bibinfo {pages}
  {1758} (\bibinfo {year} {1999})}\BibitemShut {NoStop}%
\bibitem [{\citenamefont {Perdew}\ \emph {et~al.}(1996)\citenamefont {Perdew},
  \citenamefont {Burke},\ and\ \citenamefont
  {Ernzerhof}}]{Perdew_PRLet_PBEGGA}%
  \BibitemOpen
  \bibfield  {author} {\bibinfo {author} {\bibfnamefont {J.~P.}\ \bibnamefont
  {Perdew}}, \bibinfo {author} {\bibfnamefont {K.}~\bibnamefont {Burke}}, \
  and\ \bibinfo {author} {\bibfnamefont {M.}~\bibnamefont {Ernzerhof}},\
  }\href@noop {} {\bibfield  {journal} {\bibinfo  {journal} {Physical Review
  Letters}\ }\textbf {\bibinfo {volume} {77}},\ \bibinfo {pages} {3865}
  (\bibinfo {year} {1996})}\BibitemShut {NoStop}%
\bibitem [{\citenamefont {Perdew}\ \emph {et~al.}(1997)\citenamefont {Perdew},
  \citenamefont {Burke},\ and\ \citenamefont
  {Ernzerhof}}]{Perdew_PRLet_PBEGGAErrata}%
  \BibitemOpen
  \bibfield  {author} {\bibinfo {author} {\bibfnamefont {J.~P.}\ \bibnamefont
  {Perdew}}, \bibinfo {author} {\bibfnamefont {K.}~\bibnamefont {Burke}}, \
  and\ \bibinfo {author} {\bibfnamefont {M.}~\bibnamefont {Ernzerhof}},\
  }\href@noop {} {\bibfield  {journal} {\bibinfo  {journal} {Physical Review
  Letters}\ }\textbf {\bibinfo {volume} {78}},\ \bibinfo {pages} {1396}
  (\bibinfo {year} {1997})}\BibitemShut {NoStop}%
\bibitem [{\citenamefont {Davidson}(1983)}]{davidson_methcomputmolphys_1983}%
  \BibitemOpen
  \bibfield  {author} {\bibinfo {author} {\bibfnamefont {E.~R.}\ \bibnamefont
  {Davidson}},\ }in\ \href@noop {} {\emph {\bibinfo {booktitle} {Methods in
  computational molecular physics}}},\ \bibinfo {series and number} {\bibinfo
  {series} {NATO Advanced Study Institute}\ No.\ \bibinfo {number} {113}},\
  \bibinfo {editor} {edited by\ \bibinfo {editor} {\bibfnamefont {G.~H.~F.}\
  \bibnamefont {Diercksen}}\ and\ \bibinfo {editor} {\bibfnamefont
  {S.}~\bibnamefont {Wilson}}},\ \bibinfo {organization} {NATO Scientific
  Affairs Division}\ (\bibinfo  {publisher} {D. Reidel Publishing Company},\
  \bibinfo {address} {Dordrecht, Holland},\ \bibinfo {year} {1983})\ pp.\
  \bibinfo {pages} {95--113}\BibitemShut {NoStop}%
\bibitem [{mat(2010)}]{matlab}%
  \BibitemOpen
  \href@noop {} {\emph {\bibinfo {title} {MATLAB version R2010b}}},\ \bibinfo
  {organization} {The MathWorks Inc.},\ \bibinfo {address} {Natick, MA}
  (\bibinfo {year} {2010})\BibitemShut {NoStop}%
\bibitem [{\citenamefont {Sofo}\ and\ \citenamefont
  {Mahan}(1998)}]{PhysRevB.58.15620}%
  \BibitemOpen
  \bibfield  {author} {\bibinfo {author} {\bibfnamefont {J.~O.}\ \bibnamefont
  {Sofo}}\ and\ \bibinfo {author} {\bibfnamefont {G.~D.}\ \bibnamefont
  {Mahan}},\ }\href {\doibase 10.1103/PhysRevB.58.15620} {\bibfield  {journal}
  {\bibinfo  {journal} {Physical Review B}\ }\textbf {\bibinfo {volume} {58}},\
  \bibinfo {pages} {15620} (\bibinfo {year} {1998})}\BibitemShut {NoStop}%
\bibitem [{\citenamefont {Koga}\ \emph {et~al.}(2005)\citenamefont {Koga},
  \citenamefont {Akai}, \citenamefont {Oshiro},\ and\ \citenamefont
  {Matsuura}}]{PhysRevB.71.155119}%
  \BibitemOpen
  \bibfield  {author} {\bibinfo {author} {\bibfnamefont {K.}~\bibnamefont
  {Koga}}, \bibinfo {author} {\bibfnamefont {K.}~\bibnamefont {Akai}}, \bibinfo
  {author} {\bibfnamefont {K.}~\bibnamefont {Oshiro}}, \ and\ \bibinfo {author}
  {\bibfnamefont {M.}~\bibnamefont {Matsuura}},\ }\href {\doibase
  10.1103/PhysRevB.71.155119} {\bibfield  {journal} {\bibinfo  {journal}
  {Physical Review B}\ }\textbf {\bibinfo {volume} {71}},\ \bibinfo {pages}
  {155119} (\bibinfo {year} {2005})}\BibitemShut {NoStop}%
\bibitem [{\citenamefont {Kuznetsov}\ \emph {et~al.}(2003)\citenamefont
  {Kuznetsov}, \citenamefont {Kuznetsova},\ and\ \citenamefont
  {Rowe}}]{0953-8984-15-29-315}%
  \BibitemOpen
  \bibfield  {author} {\bibinfo {author} {\bibfnamefont {V.~L.}\ \bibnamefont
  {Kuznetsov}}, \bibinfo {author} {\bibfnamefont {L.~A.}\ \bibnamefont
  {Kuznetsova}}, \ and\ \bibinfo {author} {\bibfnamefont {D.~M.}\ \bibnamefont
  {Rowe}},\ }\href {http://stacks.iop.org/0953-8984/15/i=29/a=315} {\bibfield
  {journal} {\bibinfo  {journal} {Journal of Physics: Condensed Matter}\
  }\textbf {\bibinfo {volume} {15}},\ \bibinfo {pages} {5035} (\bibinfo {year}
  {2003})}\BibitemShut {NoStop}%
\bibitem [{\citenamefont {Chaput}\ \emph {et~al.}(2005)\citenamefont {Chaput},
  \citenamefont {P{\'e}cheur}, \citenamefont {Tobola},\ and\ \citenamefont
  {Scherrer}}]{PhysRevB.72.085126}%
  \BibitemOpen
  \bibfield  {author} {\bibinfo {author} {\bibfnamefont {L.}~\bibnamefont
  {Chaput}}, \bibinfo {author} {\bibfnamefont {P.}~\bibnamefont {P{\'e}cheur}},
  \bibinfo {author} {\bibfnamefont {J.}~\bibnamefont {Tobola}}, \ and\ \bibinfo
  {author} {\bibfnamefont {H.}~\bibnamefont {Scherrer}},\ }\href {\doibase
  10.1103/PhysRevB.72.085126} {\bibfield  {journal} {\bibinfo  {journal} {Phys.
  Rev. B}\ }\textbf {\bibinfo {volume} {72}},\ \bibinfo {pages} {085126}
  (\bibinfo {year} {2005})}\BibitemShut {NoStop}%
\bibitem [{\citenamefont {Shi}\ \emph {et~al.}(2005)\citenamefont {Shi},
  \citenamefont {Zhang}, \citenamefont {Chen},\ and\ \citenamefont
  {Yang}}]{Shi_FFL_2005}%
  \BibitemOpen
  \bibfield  {author} {\bibinfo {author} {\bibfnamefont {X.}~\bibnamefont
  {Shi}}, \bibinfo {author} {\bibfnamefont {W.}~\bibnamefont {Zhang}}, \bibinfo
  {author} {\bibfnamefont {L.~D.}\ \bibnamefont {Chen}}, \ and\ \bibinfo
  {author} {\bibfnamefont {J.}~\bibnamefont {Yang}},\ }\href {\doibase
  10.1103/PhysRevLett.95.185503} {\bibfield  {journal} {\bibinfo  {journal}
  {Phys. Rev. Lett.}\ }\textbf {\bibinfo {volume} {95}},\ \bibinfo {pages}
  {185503} (\bibinfo {year} {2005})}\BibitemShut {NoStop}%
\bibitem [{\citenamefont {Mei}\ \emph {et~al.}(2006)\citenamefont {Mei},
  \citenamefont {Zhang}, \citenamefont {Chen},\ and\ \citenamefont
  {Yang}}]{PhysRevB.74.153202}%
  \BibitemOpen
  \bibfield  {author} {\bibinfo {author} {\bibfnamefont {Z.~G.}\ \bibnamefont
  {Mei}}, \bibinfo {author} {\bibfnamefont {W.}~\bibnamefont {Zhang}}, \bibinfo
  {author} {\bibfnamefont {L.~D.}\ \bibnamefont {Chen}}, \ and\ \bibinfo
  {author} {\bibfnamefont {J.}~\bibnamefont {Yang}},\ }\href {\doibase
  10.1103/PhysRevB.74.153202} {\bibfield  {journal} {\bibinfo  {journal}
  {Physical Review B}\ }\textbf {\bibinfo {volume} {74}},\ \bibinfo {pages}
  {153202} (\bibinfo {year} {2006})}\BibitemShut {NoStop}%
\bibitem [{\citenamefont {Shi}\ \emph {et~al.}(2007)\citenamefont {Shi},
  \citenamefont {Zhang}, \citenamefont {Chen}, \citenamefont {Yang},\ and\
  \citenamefont {Uher}}]{PhysRevB.75.235208}%
  \BibitemOpen
  \bibfield  {author} {\bibinfo {author} {\bibfnamefont {X.}~\bibnamefont
  {Shi}}, \bibinfo {author} {\bibfnamefont {W.}~\bibnamefont {Zhang}}, \bibinfo
  {author} {\bibfnamefont {L.~D.}\ \bibnamefont {Chen}}, \bibinfo {author}
  {\bibfnamefont {J.}~\bibnamefont {Yang}}, \ and\ \bibinfo {author}
  {\bibfnamefont {C.}~\bibnamefont {Uher}},\ }\href {\doibase
  10.1103/PhysRevB.75.235208} {\bibfield  {journal} {\bibinfo  {journal}
  {Physical Review B}\ }\textbf {\bibinfo {volume} {75}},\ \bibinfo {pages}
  {235208} (\bibinfo {year} {2007})}\BibitemShut {NoStop}%
\bibitem [{\citenamefont {Nolas}\ \emph {et~al.}(1996)\citenamefont {Nolas},
  \citenamefont {Slack}, \citenamefont {Morelli}, \citenamefont {Tritt},\ and\
  \citenamefont {Ehrlich}}]{NolasSlacketal_PRB_1996}%
  \BibitemOpen
  \bibfield  {author} {\bibinfo {author} {\bibfnamefont {G.~S.}\ \bibnamefont
  {Nolas}}, \bibinfo {author} {\bibfnamefont {G.~A.}\ \bibnamefont {Slack}},
  \bibinfo {author} {\bibfnamefont {D.~T.}\ \bibnamefont {Morelli}}, \bibinfo
  {author} {\bibfnamefont {T.~M.}\ \bibnamefont {Tritt}}, \ and\ \bibinfo
  {author} {\bibfnamefont {A.~C.}\ \bibnamefont {Ehrlich}},\ }\href
  {http://libproxy.wustl.edu/login?url=http://search.ebscohost.com/login.aspx?direct=true&db=aph&AN=7661601&site=ehost-live&scope=site}
  {\bibfield  {journal} {\bibinfo  {journal} {Journal of Applied Physics}\
  }\textbf {\bibinfo {volume} {79}},\ \bibinfo {pages} {4002} (\bibinfo {year}
  {1996})}\BibitemShut {NoStop}%
\bibitem [{\citenamefont {Caillat}\ \emph {et~al.}(1996)\citenamefont
  {Caillat}, \citenamefont {Borshchevsky},\ and\ \citenamefont
  {Fleurial}}]{Caillat_JAP_PropertiesSingleCrystallineCoSb3}%
  \BibitemOpen
  \bibfield  {author} {\bibinfo {author} {\bibfnamefont {T.}~\bibnamefont
  {Caillat}}, \bibinfo {author} {\bibfnamefont {A.}~\bibnamefont
  {Borshchevsky}}, \ and\ \bibinfo {author} {\bibfnamefont {J.-P.}\
  \bibnamefont {Fleurial}},\ }\href {\doibase 10.1063/1.363405} {\bibfield
  {journal} {\bibinfo  {journal} {Journal of Applied Physics}\ }\textbf
  {\bibinfo {volume} {80}},\ \bibinfo {pages} {4442} (\bibinfo {year}
  {1996})}\BibitemShut {NoStop}%
\end{thebibliography}

%

\clearpage

\begin{table}
\caption{\label{table_element_breakdown}Filler characteristics}
\begin{ruledtabular}
\begin{tabular}{cccc}
Filler & Size (pm) & Atomic weight (amu) & Electronegativity \\
\hline
Ag & 144 & 108 & 1.9 \\
Au & 135 & 197 & 2.4 \\
La & 195 & 139 & 1.1 \\
\end{tabular}
\end{ruledtabular}
\end{table}

\begin{table}
\caption{\label{table_symmetries}Unfilled (CoM$_3$) and filled (with X) Co-based skutterudites}
\begin{ruledtabular}
\begin{tabular}{ccc}
Filling Fraction & Chemical Formula & Highest Computed Symmetry Space Group \\
\hline
0 & CoM$_3$ & Im$\bar{3}$ \\
0.125 & X$_{0.125}$(CoM$_3$)$_4$ & Im$\bar{3}$ \\
0.25 & X$_{0.25}$(CoM$_3$)$_4$ & R$\bar{3}$ \\
0.375 & X$_{0.375}$(CoM$_3$)$_4$ & Im$\bar{3}$ \\
0.5 & X$_{0.5}$(CoM$_3$)$_4$ & Pm$\bar{3}$ \\
0.625 & X$_{0.625}$(CoM$_3$)$_4$ & Im$\bar{3}$ \\
0.75 & X$_{0.75}$(CoM$_3$)$_4$ & Cmmm \\
0.875 & X$_{0.875}$(CoM$_3$)$_4$ & Im$\bar{3}$ \\
1 & X(CoM$_3$)$_4$ & Im$\bar{3}$ \\
\end{tabular}
\end{ruledtabular}
\end{table}

\begin{threeparttable}
\caption{\label{table_lattice}Structural parameters for unfilled skutterudites.}
\begin{ruledtabular}
\begin{tabular}{cccc}
{} & Lattice Parameter (\AA) & \multicolumn{2}{c}{Void Radius (\AA)} \\
System & Experimental\tnote{a} & Experimental\tnote{a} & Computed \\
\hline
CoSb$_3$ & 9.0385 & 1.892 & 1.875 \\
CoAs$_3$ & 8.2055 & 1.825 & 1.780 \\
CoP$_3$ & 7.7073 & 1.763 & 1.768 \\
\end{tabular}
\end{ruledtabular}
\begin{tablenotes}
 \item[a]{Experimental values from Nolas et al. \citep{NolasSlacketal_PRB_1996}}
\end{tablenotes}
\end{threeparttable}

\clearpage

\end{document}